\documentclass[aps, prx, reprint, tightenlines, letterpaper, amsmath, amssymb, preprintnumbers, floatfix, longbibliography, nofootinbib,superscriptaddress]{revtex4-2}
\usepackage[ruled]{algorithm2e}

\usepackage{tabularx}
\usepackage{dsfont}
\usepackage{extarrows}
\usepackage{amsmath}
\usepackage{amssymb}
\usepackage{physics}
\usepackage{tabu}
\usepackage{bm}
\usepackage{graphicx}
\usepackage{placeins}
\usepackage{textgreek}
\usepackage{multirow}
\usepackage{makecell}
\usepackage{soul}
\usepackage{diagbox}
\usepackage{multirow}
\usepackage{listings}
\usepackage{graphicx}  
\usepackage[caption=false]{subfig}

\usepackage[normalem]{ulem}

\usepackage{bm}

\newcolumntype{Y}{>{\centering\arraybackslash}X}

\makeatletter
\renewcommand\onecolumngrid{%
\do@columngrid{one}{\@ne}%
\def\set@footnotewidth{\onecolumngrid}%
\def\footnoterule{\kern-6pt\hrule width 1.5in\kern6pt}%
}
\makeatother

\newtheorem{theorem}{Theorem}[section]

\usepackage[hypertexnames=false]{hyperref}
\hypersetup{
   colorlinks=true,       
   linkcolor=blue,          
   citecolor=blue,        
   filecolor=blue,      
   urlcolor=blue           
}
\usepackage{orcidlink}

\begin{document}

\title{Preparing thermal states of frustrated quantum spin systems using 139 qubits}

\author{Roland C. Farrell${}^{\orcidlink{0000-0001-7189-0424}}$}
\email{rolandf@caltech.edu}
\affiliation{Institute for Quantum Information and Matter, California Institute of Technology}
\affiliation{Department of Physics, California Institute of Technology}

\author{Yongtao Zhan${}^{\orcidlink{0000-0002-9314-0517}}$}
\affiliation{Institute for Quantum Information and Matter, California Institute of Technology}
\affiliation{Department of Physics, California Institute of Technology}

\author{Lucas Katschke${}^{\orcidlink{0009-0004-0334-6143}}$}
\affiliation{Department of Physics and Arnold Sommerfeld Center for Theoretical Physics (ASC), Ludwig Maximilian University of Munich, 80333 Munich, Germany}
\affiliation{Munich Center for Quantum Science and Technology (MCQST), 80799 Munich, Germany}

\author{Lode Pollet${}^{\orcidlink{0000-0002-7274-2842}}$}
\affiliation{Department of Physics and Arnold Sommerfeld Center for Theoretical Physics (ASC), Ludwig Maximilian University of Munich, 80333 Munich, Germany}
\affiliation{Munich Center for Quantum Science and Technology (MCQST), 80799 Munich, Germany}

\author{Ilan T. Rosen${}^{\orcidlink{0000-0001-8869-7364}}$}
\affiliation{IBM Quantum, IBM Research Cambridge, Cambridge, MA 02142, USA}

\author{Jad C.~Halimeh${}^{\orcidlink{0000-0002-0659-7990}}$}
\email{jad.halimeh@lmu.de}
\affiliation{Department of Physics and Arnold Sommerfeld Center for Theoretical Physics (ASC), Ludwig Maximilian University of Munich, 80333 Munich, Germany}
\affiliation{Max Planck Institute of Quantum Optics, 85748 Garching, Germany}
\affiliation{Munich Center for Quantum Science and Technology (MCQST), 80799 Munich, Germany}
\affiliation{Department of Physics, College of Science, Kyung Hee University, Seoul 02447, Republic of Korea}
\date{\today}

\begin{abstract}
\noindent
Finite-temperature properties of strongly correlated quantum matter are central to condensed matter, chemistry, and high-energy physics, yet are often inaccessible to classical methods such as quantum Monte Carlo (QMC).
Here, we investigate
dissipative thermal state preparation of frustrated spin systems using digital quantum computers. 
We focus on two paradigmatic models on the kagome lattice: the antiferromagnetic Heisenberg model (AFHM), whose finite-temperature properties are inaccessible to QMC due to a severe sign problem, and the antiferromagnetic Ising model (AFIM), which serves as a sign-problem-free benchmark.
Using IBM quantum processors, we prepare approximate thermal states of the AFIM on kagome lattices with up to $79$ spins coupled to $60$ environment qubits.
We observe the emergence of a robust steady state with an adjustable effective temperature that persists in circuits with over 1000 layers of two-qubit gates.
We further study the scalability of the dissipative protocol through classical statevector simulations of the AFIM and AFHM. 
On lattices with up to 27 sites, we find that the circuit depth to reach thermal equilibrium is independent of system size and grows at most linearly with inverse temperature.
These results establish engineered dissipation as a promising approach to finite-temperature quantum simulation of frustrated matter, and point toward regimes where quantum devices may outperform classical methods.
\end{abstract}

\maketitle

\section{Introduction}
\noindent
\noindent
Predicting finite-temperature properties of quantum matter is a central challenge across the many length scales where quantum effects are relevant: from quarks and gluons at the femtometer scale, to electronic structure and chemical bonding at angstrom scales, to collective phenomena in condensed matter and ultracold atomic systems extending over mesoscopic scales~\cite{Laermann2003LaticeQCDatfinitetemperature,Philipsen2007LatticeQCDatfinitetemperatureanddensity,Mermin1965Thermalproperties,Bonitz2020Abinitiosimulationofwarmdensematter,Bloch2008manybodyphysics,Gross2017Quantumsimulations}.
Classical computational methods generally break down when computing dynamical properties of strongly correlated quantum systems driven out of thermal equilibrium~\cite{Calabrese:2005in,M_hlbacher_2008}.
Such regimes provide compelling opportunities for quantum computers~\cite{Feynman1986,Lloyd1073}, with broad applications including charge and heat transport~\cite{bertini2021finite}, catalysis~\cite{mcardle2020quantum,von2021quantum}, and hydrodynamic response~\cite{PhysRevResearch.4.033059}. 
Beyond dynamics, there are many systems for which even computing equilibrium observables surpasses the capabilities of the best classical methods~\cite{Troyer:2004ge}. 

For many equilibrium problems, quantum Monte Carlo (QMC) methods provide the state-of-the-art numerical framework for accessing finite-temperature observables. However, QMC becomes inefficient in systems afflicted by the sign problem, where the signal-to-noise ratio decays exponentially with system size and inverse temperature $\beta$.
The sign problem afflicts many important physical systems—including certain classes of geometrically frustrated magnets~\cite{frustrated_boook} and fermions at finite density~\cite{PhysRevB.41.9301,Nagata:2021ugx}—severely limiting investigations of high-temperature superconductivity~\cite{PhysRevX.5.041041}, the QCD critical endpoint~\cite{Stephanov:2024xkn}, and dense nuclear matter~\cite{Baym:2017whm}.

Digital quantum simulation
may overcome the limitations of classical computing methods for determining both dynamical and static quantities at finite temperature.
Realizing such a quantum advantage requires a scalable protocol for preparing thermal (Gibbs) states on a quantum computer~\cite{Ding:2025ulc, jiang2024quantum, chen2025efficient, chen2023quantum,hahn2025towards,Lloyd:2025cvp}.
One promising approach is to couple the system to an engineered environment and repeatedly apply a dissipative quantum channel that satisfies approximate quantum detailed balance~\cite{chen2023quantum,chen2026overcoming}.
This process drives the system toward its Gibbs state without requiring prior knowledge of its structure.
Dissipative state preparation avoids the entropy estimation in methods that minimize the free energy~\cite{ consiglio2023variational,Li:2025rik} and does not require the deep circuits used in imaginary time evolution~\cite{motta2020determining}, phase estimation~\cite{Poulin:2009uco,Riera:2012ywi,Yung:2010pug,Temme:2009wa} or adiabatic evolution~\cite{chen2023efficient}.
As we show, these advantages of dissipative quantum algorithms, together with their inherent robustness to noise~\cite{Mi:2023evq,Brunner:2024ejl,Song:2025pwd},
enable the preparation of approximate thermal states on current quantum computers at significantly larger scales than have previously been achieved~\cite{PRXQuantum.2.010317,PhysRevA.110.012445,Granet:2025sks,Li:2025rik,Robertson:2026com}.

In this work, we prepare Gibbs states in frustrated spin models on the two-dimensional kagome lattice using the dissipative quantum algorithm introduced in Ref.~\cite{Ding:2025ulc}; see also Ref.~\cite{Lloyd:2025cvp}. 
In this method, environment qubits are coupled to the system via two-qubit Paulis and dissipation is implemented via mid-circuit qubit reset.
The protocol has rigorous bounds on the fidelity of the prepared Gibbs state (in the absence of noise), and only requires simple operations that are supported by existing quantum hardware.
The first model we study is the antiferromagnetic Heisenberg model (AFHM), which is a candidate quantum spin liquid~\cite{Balents:2010wrb} and is approximately realized in low-dimensional materials~\cite{Norman_2016}.
The low-temperature phase of the AFHM is inaccessible to QMC due to a severe QMC sign problem, and the structure of its ground state is actively debated~\cite{Yan:2010tty,PhysRevB.84.020407,PhysRevX.15.011047,Sun_2024}.
The second model is the antiferromagnetic Ising model (AFIM), which retains the salient features of geometric frustration without a QMC sign problem.
Its simple Hamiltonian makes the AFIM a natural testbed for studying frustrated magnetism on current quantum hardware,
and aspects of its low energy physics have previously been explored using D-Wave's quantum annealer~\cite{Narasimhan:2023inw}.

We analyze scalability through classical statevector simulations of dissipative thermal state preparation on kagome lattices with $12$, $18$, $24$ and $27$ sites. 
The thermalization dynamics of frustrated systems is limited by transitions that connect the large manifold of low-energy states to the rest of the spectrum.
For the AFIM, this leads to a surprising phenomenon where lower temperatures can thermalize faster than higher ones.
Nevertheless, we find evidence that the quantum runtime to reach constant fidelity density with the exact thermal state is independent of system size, and scales at most linearly with $\beta$.
The system-size independence is striking and suggests that the quantum channel is rapidly mixing~\cite{kastoryano2013quantum,wzb3-dbg9}.
If this behavior persists asymptotically, it would imply an exponential runtime advantage over QMC for the AFHM at low temperatures, indicating that engineered dissipation is an extensible approach toward thermal state preparation in regimes that are classically intractable.

We further demonstrate near-term feasibility by preparing approximate thermal states in the AFIM on kagome lattices with up to $79$ sites coupled to $60$ environment sites using IBM's quantum computers.
We observe that noise drives the system to a temperature-dependent steady state with an energy density that is higher than the corresponding noiseless value.
A steady state persists for all circuit depths explored, including up to 22 dissipative cycles with a two-qubit gate depth exceeding 1000.
A comparison of local observables to QMC reveals that the steady state is not described by a single global temperature.
Nevertheless, the local magnetization and antiferromagnetic correlations exhibit the temperature dependence and boundary sensitivity expected from QMC.
These results show that dissipative quantum algorithms can robustly capture nontrivial finite-temperature signatures even in the presence of hardware noise.
\begin{figure*}[!ht]
    \centering
    \includegraphics[width=0.9\linewidth]{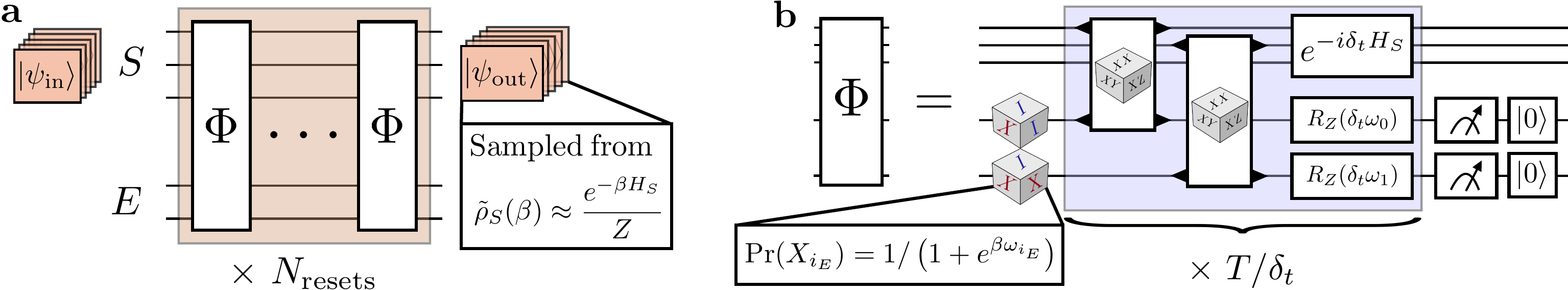}
    \caption{{\bf a} The thermal state preparation algorithm from Ref.~\cite{Ding:2025ulc}. Qubits are split into system (S) and environment (E).
    The system qubits are acted on by the quantum channel $\Phi$ a total of $N_{\text{resets}}$ times.
    This produces an ensemble of pure states $\{|\psi_{\text{out}}\rangle\}$ that are sampled from $\tilde{\rho}_S(\beta)$ which approximates the Gibbs ensemble of the system Hamiltonian $H_S$ at inverse temperature $\beta$.
    {\bf b} The quantum channel $\Phi$ begins by initializing each environment qubit to a random product state sampled from  the Gibbs ensemble of the environment Hamiltonian $H_E$.
    Next, evolution for time $T$ is implemented in discrete time steps $\delta_t$.
    The total Hamiltonian contains $H_S$, $H_E$ and a system-environment interaction component, $H_{SE}$, consisting of two-qubit Paulis in a random basis.
    After time evolution, the environment qubits are measured and reset.}
    \label{fig:overview}
\end{figure*}
%

\section{Dissipative Quantum Gibbs Sampling}
\label{sec:algOverview}
\noindent
Given a system Hamiltonian $H_S$, the Gibbs state ensemble at inverse temperature $\beta$ is $\rho_S(\beta)=e^{-\beta H_S}/Z$ with $Z={\rm Tr}(e^{-\beta H_S})$. The Gibbs ensemble is a mixed state for finite $\beta$ and can be decomposed as a convex sum over pure states $\rho_S(\beta)=\sum_i p_i |\psi_i\rangle\langle\psi_i|$, where the $p_i$ are positive probabilities that sum to one.
This decomposition is not unique~\cite{HUGHSTON199314}; a familiar one is the (spectral) ensemble of energy eigenstates with $|\psi_i\rangle=|E_i\rangle$ and $p_i=e^{-\beta E_i}/Z$. 
The Gibbs sampler considered in this work implements a quantum channel in which each trajectory (shot) prepares a pure state $|\psi_i\rangle$ with probability $p_i$ sampled from a convex decomposition of the Gibbs state.
The decomposition is determined by the dynamics of the Gibbs sampler and is not necessarily the spectral ensemble.
Thermal expectations of observables are obtained by averaging over many trajectories.
Inspired by natural thermalization processes, several works proposed preparing Gibbs states by coupling the system to an environment with a tunable temperature~\cite{shtanko2021preparing, hagan2025thermodynamic,Brunner:2024ejl}. 
The resulting dynamics between system and environment drive the system toward the target thermal state.
Although conceptually simple, efficient Gibbs sampling protocols for digital quantum computers with rigorous bounds on the convergence to a thermal ensemble did not exist until very recently~\cite{Ding:2025ulc,Lloyd:2025cvp}.

In this work, we use the quantum Gibbs sampling method from Ref.~\cite{Ding:2025ulc}.
The qubits are partitioned into $N_S$ system qubits and $N_E$ environment qubits.
A quantum channel $\Phi$ is implemented that acts on the system and has a unique steady state $\Phi[\tilde{\rho}_S(\beta)] = \tilde{\rho}_S(\beta)$.
The steady state approximates the thermal state, $||\tilde{\rho}_S(\beta)-\rho_S(\beta)||_1<\epsilon$, where $\epsilon$ is a systematically improvable error. 
This is due to the dynamics approximately satisfying a quantum detailed balance condition with respect to the Gibbs state.
The channel $\Phi$ is applied repeatedly to an arbitrary input state for a total of $N_{\text{resets}}$ iterations, producing an output state $|\psi_{\text{out}}\rangle$ in each run that is pure in absence of noise. 
Repeating this procedure many times for sufficiently large $N_{\text{resets}}$ produces an ensemble of output states that are sampled from $\tilde{\rho}_S(\beta)$, as illustrated in Fig.~\ref{fig:overview}{\bf a}.

The channel $\Phi$ has three components shown in  Fig.~\ref{fig:overview}{\bf b}.
First, the environment qubits are prepared in a product state sampled from the Gibbs ensemble of the classical environment Hamiltonian $H_E=-\sum_{i_E}\omega_{i_E}Z_{i_E}/2$.
The $\omega_{i_E}\in(0,\omega_{\text{max}}]$ are Bohr frequencies that are randomly sampled from a uniform distribution. 
Each environment qubit couples to the system via $H_{SE} = \sum_{\langle i_S,i_E\rangle} O_{i_S}\otimes X_{i_E}$ where the $O_{i_S}$ are jump operators.
The subscript $S$ and $E$ on site labels correspond to system and environment sites respectively.
These jump operators induce transitions between the energy eigenstates of $H_S$ and, unless otherwise specified, are chosen randomly from single Pauli operators, i.e., $O_{i_S} \in \{X_{i_S}, Y_{i_S}, Z_{i_S}\}$.
The joint system–environment state is then evolved for a time $T$ under the total Hamiltonian $H(t)=H_S+H_E+\alpha f(t)H_{SE}$ where $f(t)$ is a Gaussian filter function and $\alpha$ is the coupling strength.
The filter function sets the energy resolution of induced transitions.
After the time evolution, the environment qubits are measured and reset to $|0\rangle$.
More details on the dissipative algorithm are given in Methods~\ref{methods:algo}.

The number of applications of the quantum channel $\Phi$ required to reach the steady state depends on the mixing time $\tau_{\text{mix}}$.
Dissipative preparation of thermal states is formally efficient if $\tau_{\text{mix}}$ scales polynomially in $N_S$ and $\beta$.
Such polynomial mixing has been rigorously established for local Hamiltonians at high temperature~\cite{rouze2025efficient,rouze2026optimal,slezak2026polynomial}, for one-dimensional local Hamiltonians at any constant temperature~\cite{bergamaschi2025quantum}, and for weakly interacting systems of fermions~\cite{tong2025fast,vsmid2025polynomial} and bosons~\cite{Smid:2025gks}.
However, these results are restricted to regimes that are classically tractable \cite{bakshi2024high,PhysRevX.11.011047,chen2025convergence}, and do not extend to regimes believed to be classically intractable, such as frustrated systems at low temperatures.

\section{Classical simulations of Gibbs sampling in frustrated systems}
\label{sec:classicalsims}
\noindent
In this section, the quantum Gibbs sampling algorithm is implemented using a classical statevector simulator.
We study the AFIM and AFHM on kagome lattices, and estimate how $\tau_{\text{mix}}$ scales with $N_S$ and $\beta$.
We refer to the Methods~\ref{methods:FrustratedPhysics} section for an overview of the low-energy properties of both models.

\subsection{Setup}
\noindent 
The AFIM and AFHM Hamiltonians are
\begin{align}
    &H_{\text{AFIM}} \ = \  \sum_{\langle {i_S},{j_S} \rangle}Z_{i_S} Z_{j_S} \ + \  \sum_{i_S} (g_x X_{i_S} + g_z Z_{i_S}) \ , \nonumber \\ 
    &H_{\text{AFHM}} \ = \ \sum_{\langle i_S,j_S \rangle}\left ( X_{i_S} X_{j_S}+Y_{i_S} Y_{j_S}+Z_{i_S} Z_{j_S} \right )\ ,
    \label{Eq:HKagome}
\end{align}
where $\langle i_S, j_S\rangle$ represents nearest neighbors on the kagome lattice and $g_x$, $g_z$ are the transverse and longitudinal field strengths.
The kagome lattice has a three-site unit cell; with $L_x \times L_y$ unit cells, the total number of sites is $N_S = 3 L_x L_y$.
We consider lattices with periodic boundary conditions (PBCs) and $(L_x,L_y) = (2,2),(2,3),(2,4),(3,3)$ corresponding to $N_S=12,18,24,27$. 
The lattice geometries are shown in Appendix~\ref{app:csim}.
\begin{figure*}
    \centering
    \includegraphics[width=\linewidth]{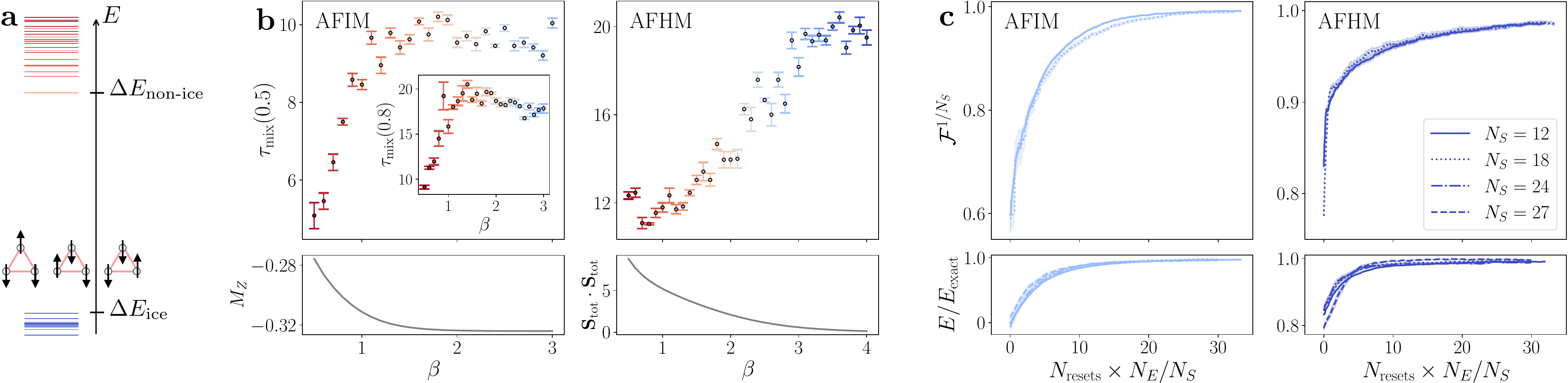}
    \caption{{\bf a} The low energy spectrum of the AFIM for $g_z=2,g_x=0.5$.
    {\bf b} The mixing time, $\tau_{\text{mix}}({\cal F}_{\text{thresh}})$ for $N_S=12$, a range of $\beta$, and in the AFIM (left) and AFHM (right).
    The inset gives the mixing time for a larger fidelity threshold in the AFIM showing a local maxima at the ice crossover, $\beta_c\approx1.4$.
    The bottom plots show the $\beta$-dependence of the average magnetization $M_Z$ in the AFIM and the square of the total spin ${\bf S}_{\text{tot}}\cdot{\bf S}_{\text{tot}}$ in the AFHM.
    {\bf c} The convergence of the quantum Gibbs samplers for increasing lattice sizes.
    The upper plots show the fidelity in the AFIM with $\beta=3$ (left) and AFHM with $\beta = 4$ (right).
    The  bottom panels show the ratio of the measured energy $E$ to the exact energy $E_{\text{exact}}$.
    The parameters for {\bf b} and {\bf c} are given in the caption of Fig.~\ref{fig:IsingHeisFid}. The results are obtained from a noise-free numerical simulation.
    }
    \label{fig:mix_scaling}
\end{figure*}

In addition to discrete lattice symmetries, the AFHM has a global $SU(2)$ symmetry generated by
${\bf S}_{\text{tot}} = \sum_{i_S} {\bm \sigma}_{i_S}$
where ${\bm \sigma}_{i_S}=\{X_{i_S},Y_{i_S},Z_{i_S}\}$.
Finite-size studies indicate that the ground state and low-energy excitations are in the singlet ($S = 0$) representation~\cite{Yan:2010tty}.
This symmetry forbids the $S=1$ jump operators $O_{i_S}\in\{X_{i_S},Y_{i_S},Z_{i_S}\}$ from inducing transitions within the low-energy manifold of singlets.
To speed up thermalization, we also include $S=0$ jump operators of the form $O_{i_S,j_S}=\left ( X_{i_S}X_{j_S}+Y_{i_S}Y_{j_S}+Z_{i_S}Z_{j_S} \right )/3$ for the AFHM, as discussed in Methods~\ref{methods:jump}.

The low-energy states of the AFIM with small transverse field satisfy the ``ice rule" where two out of the three 
bonds on each triangle are antiferromagnetic~\cite{Moessner_2000,Moessner_2001,Narasimhan:2023inw}.
For $g_z>0$, the longitudinal field selects the ice configurations with $M_Z = \sum_i \langle Z_i\rangle /N_s= -1/3$ as the low-energy sector.
In this work, we choose parameters $g_x=0.5$ and $g_z=2$, with an ice to non-ice crossover temperature of $\beta > \beta_c \approx 1.4$ for our system size. 
The sign of $g_x$ is a gauge choice.
The low-energy spectrum for $N_S=24$ is shown in Fig.~\ref{fig:mix_scaling}{\bf a}, along with the three degenerate spin configurations on a triangle. 
At these parameters, there are many (exponential in $N_S$) states packed into a small energy interval $\Delta E_{\text{ice}}$ around the ground state. 
To lowest order in $g_x$, these states are superpositions of the classical spin-ice configurations, all of which have $M_Z=-1/3$. 
For this study, we do not consider temperatures that would resolve the individual ice-states, i.e. we keep $\beta \Delta E_{\rm ice} \ll 1$.
In the thermodynamic limit, degenerate perturbation theory predicts $\Delta E_{\rm ice} \approx 0.004$~\cite{Kreissel2026}.
However, in the finite-size lattices we consider, additional loops can be constructed across the periodic boundaries leading to $\Delta E_{\rm ice} \approx 0.1$. 
The remainder of the spectrum is separated from the ice manifold by a larger energy gap $\Delta E_{\text{non-ice}}\approx 1.7$.
The separation of scales gives a temperature window of $1.2 \lesssim \beta \lesssim 5$ where the physics is approximately unchanging.

\subsection{Results}
\noindent
The mixing time is defined to be the number of resets required to reach a threshold fidelity ${\cal F}_{\text{thresh}}$ with the thermal density matrix,
\[
\tau_{\text{mix}}({\cal F}_{\text{thresh}}) \ = \ N_{\text{resets}}({\cal F} > {\cal F}_{\text{thresh}} ) \times N_E/N_S \ .
\]
Rescaling by $N_E/N_S$ removes the dependence on $N_E$ as explained in Appendix~\ref{app:nQ_convergence}, and all classical simulations are performed with $N_E=1$.
The scaling of the mixing time with $\beta$ for $N_S=12$ is shown in Fig.~\ref{fig:mix_scaling}{\bf b}.
The initial state is a random product state ($\beta=0$) and, therefore, convergence is fastest for low $\beta$.
For the AFHM, the mixing time increases approximately linearly with $\beta$.
Notably, there are no features at the crossover to a singlet-dominated thermal ensemble, as quantified by $\text{Tr}\left (\rho_S(\beta)\, {\bm S}_{\text{tot}}\cdot {\bm S}_{\text{tot}} \right )$ approaching to zero, shown in the bottom right plot of Fig.~\ref{fig:mix_scaling}{\bf b}.
This crossover marks the onset of the low-temperature regime where the frustrated manifold of singlets dominates and a quantum spin liquid may emerge~\cite{1998EPJB,2011JPSJ}.
The smooth behavior of $\tau_{\text{mix}}$ indicates that the combination of $S=1$ and $S=0$ jump operators is able to efficiently thermalize across a range of temperatures.

The mixing time of the AFIM exhibits qualitatively different behavior.
For ${\cal F}_{\text{thresh}} = 0.5$, the mixing time increases for $\beta<\beta_c\approx1.4$ and then plateaus for $\beta_c<\beta<3.0$.
For temperatures in the range $\beta_c < \beta < 3.0$, the Gibbs state is well approximated by an equal-weight ensemble over the eigenstates in the ice manifold, since $\beta \Delta E_{\text{ice}} \ll 1$.
The thermal ensemble therefore changes very little over this range of temperatures and the mixing time is approximately constant.
In Appendix~\ref{app:csim}, it is shown that the mixing time increases again for $\beta \gtrsim 1/\Delta E_{\text{ice}}$ as the eigenstates in the ice manifold develop different Boltzmann weights.

Raising the fidelity threshold produces a local maximum in the mixing time at $\beta_c\approx1.4$ as shown for ${\cal F}_{\text{thresh}}=0.8$ in the inset of Fig.~\ref{fig:mix_scaling}{\bf b}.
This indicates that thermalization can be faster for $\beta > \beta_c$ than at $\beta = \beta_c$. 
This behavior is surprising because the initial state is at infinite temperature and, naively, one might expect the system to pass through $\beta_c$ when cooling to a target temperature $\beta > \beta_c$.
Under this intuition, cooling below the crossover temperature would inherit any slowdown associated with cooling to $\beta_c$.
However, this reasoning incorrectly assumes that the system is in thermal equilibrium throughout the dissipative process.
In quantum Gibbs sampling, faster thermalization can be achieved by keeping the system out of equilibrium until the fixed point is reached.
Methods~\ref{methods:jump} provides an explanation for this mixing-time slowdown based on the quantum detailed balance condition and the jump operator connectivity between energy eigenstates.
Additional information on the rate of convergence to the steady state and the methods used to compute the fidelity and error bars are provided in Appendix~\ref{app:fid}.
\begin{figure*}
    \centering
    \includegraphics[width=\linewidth]{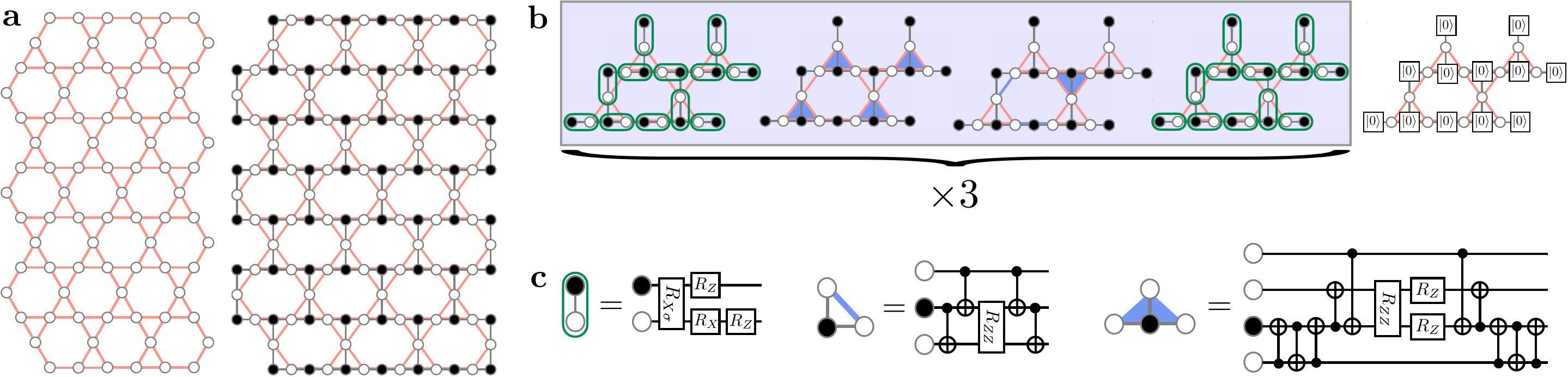}
    \caption{{\bf a} Embedding of a kagome lattice (left) onto the heavy-hex connectivity of IBM's {\tt Heron r3} processors (right).
    The heavy-hex connectivity is represented by the gray links.
    The sites (links) of the kagome lattice are defined by the white colored qubits (pink links). 
    The environment qubits (black colored qubits) are interspersed throughout the lattice.
    {\bf b} The quantum circuits used in one application of the dissipative quantum channel $\Phi$ for $(N_S,N_E)=(12,12)$ on {\tt ibm\_boston}.
    The circuits highlighted in purple correspond to one second order Trotter step, which is repeated $T/\delta_t=3$ times.
    After time evolution, the environment qubits are measured and reset to $|0\rangle$.
    {\bf c} The definition of the circuit elements in {\bf b}.
    The green chain link corresponds to the single qubit rotations as well as the system-environment coupling, the blue bond implements a next-to-nearest neighbor $R_{ZZ}(\theta)$ and the blue triangle implements $R_{ZZ}(\theta)$ between all three pairs of its vertices.
    The ordering of gates is reversed in the second application of the green chain link, as needed for second-order Trotterization.}
    \label{fig:ibm_circuits}
\end{figure*}

Next, we analyze how the mixing time scales with $N_S$ at fixed but large $\beta$.
To speed up convergence to the steady state in the AFHM, and reduce classical computing overhead, the system is initialized in a low-energy 
$S=0$ state  $|\psi_{\text{in}}\rangle  =  \left [(|01\rangle - |10\rangle)/\sqrt{2}\right ]^{\otimes N_S/2} $.
This state consists of singlet dimers on nearest-neighbor sites and has energy $\langle\psi_{\text{in}}| H_{\text{AFHM}}|\psi_{\text{in}}\rangle = -3N_S/2$.
For $N_S=27$, one spin is left unpaired, resulting in a slightly higher initial energy density.
For large $\beta$, the steady state is reached much faster than starting from $\beta=0$.
Furthermore, $|\psi_{\text{in}}\rangle$
can be prepared with a single layer of two-qubit gates~\cite{Lee:2026xfm}.

The rate of convergence to the steady state for $N_S=12, 18, 24,27$ are compared in Fig.~\ref{fig:mix_scaling}{\bf c}.
To enable a meaningful comparison across system sizes, the fidelity is rescaled as ${\cal F}^{1/N_S}$. 
For $N_S=24,27$, only the energy is computed due to memory constraints. 
Both the rescaled fidelity (upper panels) and the energy (lower panels) exhibit almost complete system size independence.
This suggests that both the AFIM and AFHM are rapid mixing for $N_E=N_S$, characterized by local observables reaching a fixed error within a constant number of resets. 
Combined with the observed (sub)-linear scaling of the mixing time with $\beta$, our results provide evidence that thermal states can be prepared very efficiently at scale. 

\section{Quantum simulations of Gibbs sampling in the Ising model on a kagome lattice}
\label{sec:qsims}
\noindent
In this section, we prepare approximate Gibbs states of the AFIM on kagome lattices using IBM's quantum computers.
The AFIM is chosen over the AFHM as its Hamiltonian time evolution can be implemented using approximately $2.5\times$ fewer two-qubit gates, see Appendix~\ref{app:qcircs}.
IBM's {\tt heron R3} quantum computers feature 156 superconducting qubits with heavy-hex connectivity.
The kagome lattice can be embedded into heavy-hex in a way that naturally hosts environment qubits using the lattice-to-qubit mapping from Ref.~\cite{Shinjo:2025hnu} shown in Fig.~\ref{fig:ibm_circuits}{\bf a}.
The transverse and longitudinal field strengths are set to $g_x = 0.5$ and $g_z = 2$, placing the low-temperature regime in the frustrated ice phase.

To prepare the highest quality thermal state, we want to accurately implement the quantum channel $\Phi$ while minimizing the number of two-qubit gates. 
Informed by classical simulations, we use $T/\delta_t=3$ second order Trotter steps of size $\delta_t=0.25$ during each application of $\Phi$.
Examples of the corresponding quantum circuits for $N_S=12$ and $N_E=12$ are shown in Fig.~\ref{fig:ibm_circuits}{\bf b} and {\bf c}.
The time evolution circuits utilize swap gates and the nested $R_{ZZ}(\theta)$ construction from Ref.~\cite{Farrell:2024fit} to minimize the two-qubit gate depth when mapped to heavy-hex.
In total, each reset cycle has a two-qubit gate depth of 55. 
More details on the circuit construction and parameters chosen are provided in Methods~\ref{methods:qsim}.
To suppress hardware errors we apply zero-noise extrapolation (ZNE) to address coherent gate noise and leakage detection to discard runs in which qubits leave the computational subspace.
Complete details of the error mitigation methods are provided in Appendix~\ref{app:errormitigation}.
\begin{figure*}
    \centering
    \includegraphics[width = \linewidth]{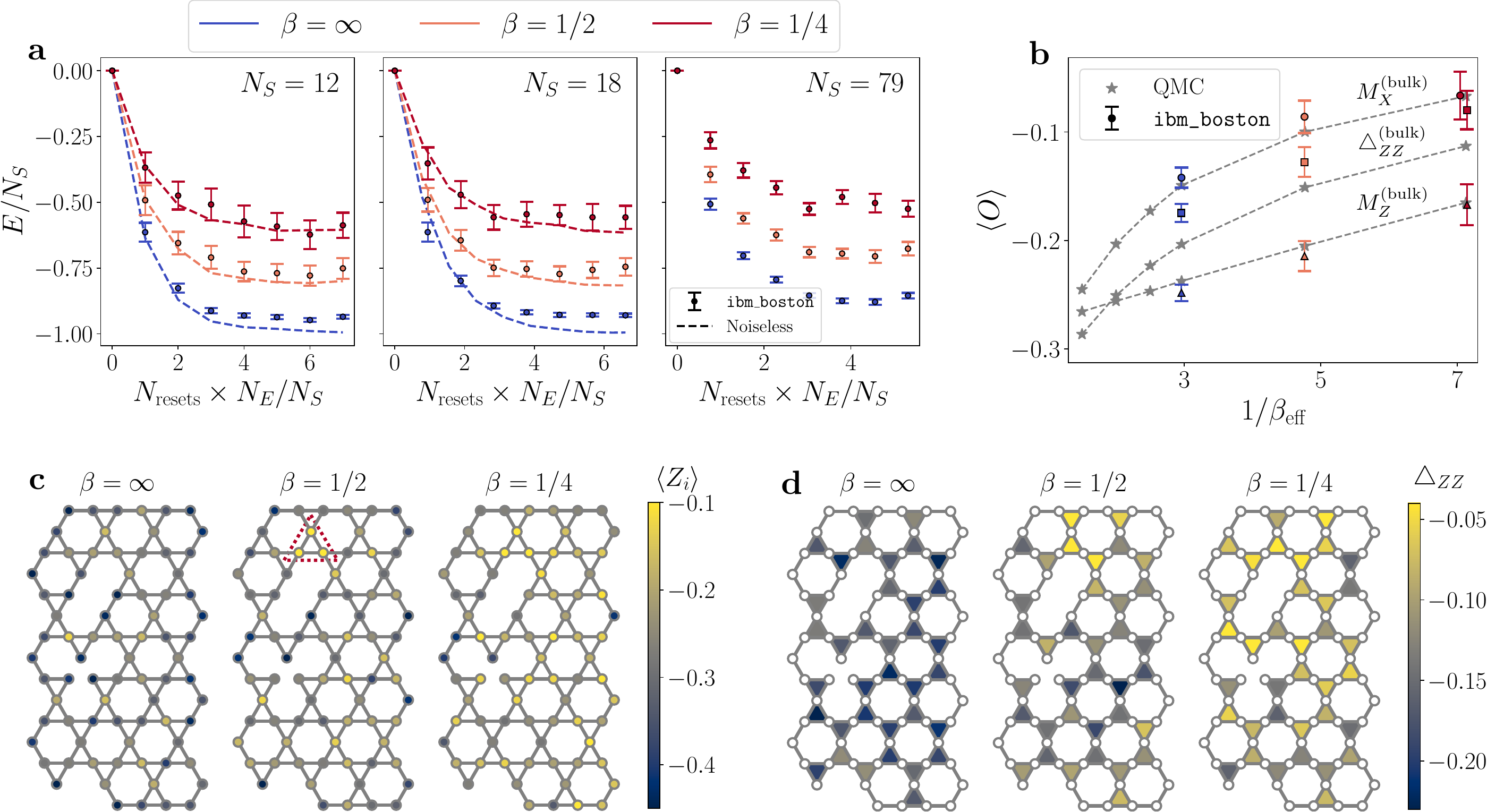}
    \caption{{\bf a} Energy density measured on {\tt ibm\_boston} for $\beta = \infty,1/2,1/4$ and systems sizes $N_S = 12,18,79$ coupled to $N_E=12,17,60$ environment sites.
    The data for $N_S=12,18$ is compared to noiseless statevector simulations with environment temperature $\beta$ (dashed).
    The x-axis is rescaled to compare the results across different $N_E/N_S$ as explained in Appendix~\ref{app:nQ_convergence}.
    The spatial distribution of magnetization {\bf c} and $\triangle_{ZZ}$ {\bf d} measured in the approximate thermal states for $N_S=79$. 
    The spins in {\bf c} that are outlined in red surround a hot environment qubit with a large reset error rate.
    Bulk averages for $N_S=79$ are compared in {\bf b} to QMC predictions at an effective temperature $\beta_{\text{eff}}$ that reproduces the steady state energy.}
    \label{fig:IBM_results_Energy_nE}
\end{figure*}
%

\subsection{Lattice cooling dynamics}
\noindent
Approximate thermal states are prepared on {\tt ibm\_boston} with $(N_S,N_E)=(12,12),(18,17),(79,60)$ and inverse bath temperatures $\beta=\infty,1/2,1/4$.
We first examine the energy densities in Fig.~\ref{fig:IBM_results_Energy_nE}{\bf a}. 
The parameter choices $(\alpha$, $T$, and $\delta_t)$  have been optimized to reduce the two-qubit gate count at the expense of a systematic deviation from the exact thermal state.
This deviation is quantified in Table~\ref{tab:energies} which reports the energy densities obtained from QMC, noise-free statevector simulations and {\tt ibm\_boston}.

Initially, at $N_{\text{resets}}=0$, the energy is zero because random product states ($\beta=0$) are initialized on the system qubits. 
Each reset cycle lowers the energy by an amount that decreases as the system converges to a steady state.
The results for $N_S=12,18$ follow noiseless expectations for the first few reset cycles, but then deviate toward higher energies, with the deviation becoming more pronounced at lower temperatures.\footnote{The error bars combine statistical and ZNE fit uncertainties.
The classical simulations for $N_S=18$ are performed with $N_E=14$ due to memory constraints.
}
The energy densities plateau for $N_{\text{resets}}\geq 4$, with further reset cycles leaving the energy unchanged within error bars.
The energy decreases as the temperature is lowered, as expected, and the plateau suggests that the system is reaching a $\beta$-dependent steady state in the presence of device noise.
In Appendix~\ref{app:noiseEffect}, it is shown that a steady state persists out to $N_{\text{resets}} = 22$ with two-qubit gate depths exceeding 1000.
Similar steady-state behavior on a quantum computer has also been observed in dissipative ground state preparation~\cite{Mi:2023evq,Song:2025pwd}.

The steady state reached in the experiments depends on the device noise.
One source of noise comes from gate errors, predominantly from the two-qubit gates.
In unitary circuits, after Pauli twirling, the noise is well described by a stochastic Pauli channel $\Phi_{\text{Pauli}}$~\cite{Wallman:2015uzh,Berg:2022ugn,Chen:2025cfe}.
This channel drives all observables to zero, and its steady state is maximally mixed.
The Gibbs sampling circuits, see Fig.~\ref{fig:ibm_circuits}{\bf b}, are not unitary, as they have mid-circuit qubit resets that remove memory in the environment from prior cycles.
They apply a quantum channel $\Phi$ that drives the system toward the approximate thermal state $\tilde{\rho}_S(\beta)$.
The Pauli noise and dissipation together realize a combined quantum channel with a new steady state $\rho_{\text{s.s.}}$.
The magnitude of the perturbation to the steady state can be calculated under the simplifying assumption of weak global depolarizing noise that acts at the end of each reset cycle. 
The result is
\begin{align}
||\rho_{\text{s.s.}} - \tilde{\rho}_S(\beta)||_1 \ \lesssim \ (p\, \tau_{\text{mix}}||\tilde{\rho}_S(\beta)-\mathds{1}/2^{N_S} ||_1) \ ,
\label{eq:noisePerturb}
\end{align}
where $p$ is noise strength. 
This bound is derived in Appendix~\ref{app:noiseEffect} along with an analysis for a more realistic noise model that shows a similar scaling for the error in local observables.
Although the actual device noise is more complex, the results from {\tt ibm\_boston} are consistent with the scaling in Eq.~\eqref{eq:noisePerturb}.
Taking $\beta\to0$ sends the system to the maximally mixed state, decreasing both $||\tilde{\rho}_S(\beta)-\mathds{1}/2^{N_S} ||_1$ and $\tau_{\text{mix}}$, and hence the sensitivity to noise.
This is reflected in Fig.~\ref{fig:IBM_results_Energy_nE}{\bf a}, where there is improved agreement between the noiseless simulations and {\tt ibm\_boston} at lower values of $\beta$.
Equation~\eqref{eq:noisePerturb} also predicts linear scaling with $\tau_{\text{mix}}$, which is supported by additional experiments that vary $N_E$ presented in Appendix~\ref{app:noiseEffect}.

A secondary source of noise comes from imperfect mid-circuit resets.
Instead of preparing $|0\rangle$, there is a probability of incorrectly resetting the environment qubits to $|1\rangle$. 
This shifts the environment to a lower inverse-temperature $\beta^*<\beta$.
Additional calibration experiments show that this effect is small for $N_S=12$ and $N_S=18$, but leads to two ``hot'' qubits on the top of the $N_S=79$ lattice.
These hot qubits generate spatial inhomogeneities in observables as discussed in the next subsection.
Further details on the determination of $\beta^*$ are provided in Appendix~\ref{app:errormitigation}.

Having benchmarked the protocol against classical simulations for $N_S = 12, 18$, we now turn to $N_S = 79, N_E = 60$.
This scale is far beyond the reach of exact and, to the best of our knowledge, even approximate classical simulations of the Gibbs sampling protocol. 
Without a noiseless benchmark, we compare the measured observables directly to QMC.
The $N_S=79$ lattice has several defects to avoid links with large two-qubit gate errors, see Appendix~\ref{app:errormitigation}.
This irregular geometry poses a challenge for QMC that we address with a custom cluster update that is described in Appendix~\ref{app:QMC}.
The energies measured on {\tt ibm\_boston} for $N_S=79$ are shown in Fig.~\ref{fig:IBM_results_Energy_nE}{\bf a} and compared to QMC in Table~\ref{tab:energies}.
Whereas the QMC energy densities barely change across system sizes, the energy densities measured on {\tt ibm\_boston} for $N_S=79$ are systematically higher across all $\beta$.
This increase may be due to the reduced  environment density, $N_E/N_S=60/79=0.76$, that increases the mixing time and hence the sensitivity to noise through Eq.~\eqref{eq:noisePerturb}.

\subsection{Characterizing the emergent steady state}
\noindent
Measurements of individual observables that contribute to the energy help to characterize the steady state.
Consider the magnetization per site $\langle Z_i\rangle$ and the connected $ZZ$ correlations on a triangle, 
\begin{align}
\triangle_{ZZ}\ = \ \frac{1}{3}\sum_{i_S,j_S\in\triangle}\left ( \langle Z_{i_S} Z_{j_S} \rangle-\langle Z_{i_S}\rangle\langle Z_{j_S} \rangle \right )\ ,
\label{eq:TriZZ}
\end{align}
where the sum is over the three nearest-neighbor pairs forming a triangle.
In the low-temperature and thermodynamic limit, QMC simulations predict $\langle Z_i\rangle \to -0.32$ and $\triangle_{ZZ}\to-0.41$, see Appendix~\ref{app:QMC}.
The large magnitude of $\triangle_{ZZ}$ is indicative of the thermally occupied states being superpositions over the frustrated ice manifold.
At higher temperatures, these quantities converge to zero, with $\triangle_{ZZ}$ being more sensitive to thermal fluctuations as it is a measure of geometric frustration.

These observables are evaluated in the steady state at $N_{\text{resets}}=7$ and displayed in Fig.~\ref{fig:IBM_results_Energy_nE}{\bf c}, {\bf d} for $N_S=79$.
The magnetization at $\beta=\infty$ shows that spins near the boundaries, including those close to the lattice defects, are more polarized than spins in the bulk.
This behavior is also observed in QMC and is due to the competition between the longitudinal field, which tends to polarize the spins, and the antiferromagnetic couplings that favor neighboring spins to be anti-aligned.
Boundary spins have fewer antiferromagnetic couplings so are more polarized at lower temperatures.
Increasing the temperature from left to right in {\bf c} shows progressive demagnetization and increased spatial homogeneity, consistent with trends observed in QMC.
Compared to the magnetization, the magnitude of the connected $\triangle_{ZZ}$ correlator in {\bf d} decreases more rapidly with increasing temperature.
Two ``hot’’ environment qubits located near the top of the lattice locally suppress the magnetic polarization and reduce frustration in the surrounding system spins.
This effect is most obvious at $\beta=0.5$, where the spins outlined by the red triangle in Fig.~\ref{fig:IBM_results_Energy_nE}{\bf c} are adjacent to an environment qubit with effective inverse temperature $\beta^*=\{1.25,0.36,0.2\}$ for $\beta=\{\infty,0.5,0.25\}$.
The data for Fig.~\ref{fig:IBM_results_Energy_nE}{\bf c}, {\bf d} with error bars is provided in Appendix~\ref{app:errormitigation}.
\begin{table}[t]
\centering
\renewcommand{\arraystretch}{1.4}
\begin{tabular}{|c|c|c||c|c|c|c||} 
\hline
$N_S$  & $\beta$  & $\beta_{\text{eff}}$  & $E_{\text{s.s.}}^{\text{(QC)}}/N_S$ & $E^{\text{(Noiseless)}}_{\text{s.s.}}/N_S$& $E_{\text{QMC}}/N_S$  \\ 
\hline\hline
12 & $\infty$ &0.41(1)  & -0.94(1)  & -0.99 & -1.40\\ \hline
 & $1/2$  &0.25(2) &  -0.75(4) & -0.80  & -1.02\\ \hline
 & $1/4$  & 0.17(2)&  -0.59(5) &  -0.60 & -0.75\\ \hline
18 &  $\infty $ & 0.40(1)  &-0.93(1) &  -1.00 & -1.40\\ \hline
  &  $1/2$ & 0.25(2) & -0.75(3)  & -0.82 & -1.02 \\ \hline
  &  $1/4$ & 0.15(2) &  -0.56(4) & -0.61  & -0.75 \\ \hline
79 & $\infty$ & 0.34(1) &-0.86(1) & -- & -1.37(1) \\ \hline
  & $1/2$ & 0.21(1) & -0.68(3)  & -- & -1.00 \\ \hline
  & $1/4$ & 0.14(1)   & -0.53(3) & --& -0.75\\ \hline
\end{tabular}
\renewcommand{\arraystretch}{1.0}
\caption{For a given $N_S$ (column 1) and $\beta$ (column 2),
the energy density obtained from {\tt ibm\_boston}, noiseless statevector simulations and QMC are given in columns 4, 5 and 6 respectively.
The energy densities in columns 4 and 5 are evaluated at $N_{\text{resets}}=7$. 
The energy of a thermal state at inverse-temperature $\beta_{\text{eff}}$ (column 3) is equal to $E_{\text{s.s.}}^{\text{(QC)}}$.}
\label{tab:energies}
\end{table}

Next, we test the extent to which the steady state reached on {\tt ibm\_boston} is thermal by determining an effective temperature $\beta_{\text{eff}}$ that reproduces the steady state energy $E_{\text{s.s.}}$.
The effective temperature solves
\begin{align}
\text{Tr}\left [H_S\,\rho_S(\beta_{\text{eff}}) \right ] \ = \ E_{\text{s.s.}}(\beta) \ ,
\label{eq:Beff}
\end{align}
where the left-hand side is evaluated from QMC using a bisection search over temperatures~\cite{Katschke:2026nmj}.
Observables evaluated at $\beta_{\text{eff}}$ are then compared to those measured on {\tt ibm\_boston}.
If the steady state deviates strongly from a thermal distribution, agreement at the level of energy alone would not imply agreement for other observables.
We consider three bulk quantities: the average $Z$-magnetization $M_Z^{(\text{bulk})}$, $X$-magnetization $M_X^{(\text{bulk})}$ and the average connected $ZZ$ correlations around a triangle, $\triangle_{ZZ}^{(\text{bulk})}$.
The average is performed over bulk sites, defined as those with four neighboring links.

The comparison between QMC and {\tt ibm\_boston} is shown in Fig.~\ref{fig:IBM_results_Energy_nE}{\bf b}.
The single-site observables are in good agreement with QMC across the full temperature range, while the $\triangle_{ZZ}$ correlations are consistently smaller in magnitude.
This implies that the bulk energy density is higher (less negative) than the average energy density, since $E_{\text{s.s.}}$ receives contributions from all three observables.
This may be due to the chosen qubit layout having a higher density of environment qubits on the boundary than in the bulk (see Appendix~\ref{app:bigKagome}).
A denser environment mixes faster, decreasing noise sensitivity through Eq.~\eqref{eq:noisePerturb}, and leading to a lower effective temperature near the boundary than in the bulk.

\section{Outlook}
\noindent
A key takeaway from this work is that increasing the density of environment qubits not only speeds up thermalization, but also enhances the robustness of the protocol to hardware noise. 
It is therefore preferable for the environment to scale extensively with system size, ideally $N_E=N_S$.
The evidence from our classical simulations for rapid mixing in frustrated systems is surprising, and suggests that dissipative thermalization with system-size-independent circuit depth may be more widespread than expected.

Several important limitations remain. Our mixing-time analysis was restricted by classical compute resources to modest-sized kagome lattices.
Extending it to larger systems will be crucial for for assessing whether dissipative thermalization remains efficient at scale.
Reaching emergent quantum spin-liquid phases will require understanding how jump operators
couple to many-body resonances, and how long-range entanglement and correlations affect the mixing time.
As Gibbs sampling progresses toward regimes inaccessible to classical numerics, it will become necessary to develop diagnostics capable of certifying that the steady state is thermal.
A possible approach is to verify that response functions satisfy the constraints coming from thermal equilibrium, such as fluctuation-dissipation relations.
On the hardware side, the lowest effective temperature we reach is $\beta_{\text{eff}} = 0.41(1)$, constrained by the number of Trotter steps that can be executed in a reset cycle without significant noise accumulation. 
The achievable temperatures will decrease with continued improvements in qubit connectivity and the fidelity of two-qubit gates. 
Ultimately, achieving global thermal equilibrium at extremely low temperatures will benefit from fault-tolerant quantum computers.

A more immediate target for quantum utility is the simulation of dynamics in systems perturbed from these approximate thermal states. 
This regime remains challenging for classical methods even at relatively high temperatures, and the states we prepare could already serve as a useful resource. 
The scalability of dissipative state preparation demonstrated here opens a path beyond equilibrium physics, such as the finite-temperature dynamics of frustrated magnets, fermionic systems, and lattice gauge theories.

\footnotesize
\begin{acknowledgements}
\noindent
Code for performing statevector simulations of the dissipative Gibbs sampler is available at
\url{https://github.com/Roland-Farrell/dissipative_Gibbs_sampling}.
We would like to thank Anthony Chen, Senrui Chen, Abhinav Kandala, Bibek Pokharel, and John Preskill for helpful discussions.
R.F.~acknowledges support from the U.S.~Department of Energy QuantISED program through the theory consortium “Intersections of QIS and Theoretical Particle Physics” at Fermilab, from the U.S.~Department of Energy, Office of Science, Accelerated Research in Quantum Computing, Quantum Utility through Advanced Computational Quantum Algorithms (QUACQ), and from the Institute for Quantum Information and Matter, an NSF Physics Frontiers Center
(PHY-2317110). 
R.F.~additionally acknowledges support from a Burke Institute prize fellowship. Y.Z. also acknowledges support from the National
Science Foundation, grant no. PHY-2317110 and from Quantum Systems Accelerator. This research used resources of the Oak Ridge Leadership Computing Facility, which is a DOE Office of Science User Facility supported under Contract DE-AC05-00OR22725.
The computations presented in this work were conducted in the Resnick High Performance Computing Center,
a facility supported by the Resnick Sustainability Institute at Caltech.
RF acknowledges the use of IBM Quantum Credits for this work. 
This research used resources of the National Energy Research
Scientific Computing Center, a DOE Office of Science User Facility
supported by the Office of Science of the U.S.~Department of Energy
under Contract No.~DE-AC02-05CH11231 using NERSC award
NERSC DDR-ERCAP0034353. 
The project/research is part of the Munich Quantum Valley, which is supported by the Bavarian state government with funds from the Hightech Agenda Bayern Plus.
L.K.,~L.P.,~and J.C.H.~acknowledge funding by the Deutsche Forschungsgemeinschaft (DFG, German Research Foundation) under Germany’s Excellence Strategy – EXC-2111 – 390814868. L.K.~and J.C.H.~acknowledge funding by the Max Planck Society and the European Research Council (ERC) under the European Union’s Horizon Europe research and innovation program (Grant Agreement No.~101165667)—ERC Starting Grant QuSiGauge. Views and opinions expressed are, however, those of the author(s) only and do not necessarily reflect those of the European Union or the European Research Council Executive Agency. Neither the European Union nor the granting authority can be held responsible for them. 
\end{acknowledgements}
\normalsize

\clearpage

\section*{Methods}
\label{sec:methods}
\setcounter{section}{0}
\renewcommand\thesection{}
\renewcommand\thesubsection{\Alph{subsection}}
\noindent 

\subsection{Dissipative quantum Gibbs sampling}
\label{methods:algo}
\noindent
The dissipative quantum algorithm for Gibbs sampling from Ref.~\cite{Ding:2025ulc} is given in Algorithm~\ref{alg:algobox}.
It involves repeated application of the quantum channel $\Phi$ that fixes an approximate Gibbs ensemble, as shown in Fig.~\ref{fig:overview}{\bf a}.
Here, we provide a complete description of the algorithm and give guidelines for choosing the parameters.
We set $\hbar=1$ throughout.

First, the environment qubits are prepared in a product state sampled from,
\begin{align}
\rho_E(\beta) \ = \ e^{-\beta H_E} / {\rm Tr}(e^{-\beta H_E}) \ .
\label{eq:HE}
\end{align}
This is the Gibbs state of the classical environment Hamiltonian,
\begin{align}
H_E \  = \  -\frac{1}{2}\sum_{i_E =0}^{N_E-1} \omega_{i_E} Z_{i_E} \ ,
\end{align}
where the $\omega_{i_E}$ are Bohr frequencies that set the spacing between energy levels. 
To speed up the rate of thermalization, these energy spacings should coincide with energy differences of $H_S$ to resonantly drive transitions between eigenstates.
In general, the energy levels of the system are unknown, and the Bohr frequencies are sampled from a uniform random distribution $\omega_{i_E} \in (0,\omega_{\text{max}}]$ at each reset cycle.
The choice of $\omega_{\text{max}}$ is independent of system size and should be roughly the size of the largest energy jump, $\omega_{\text{max}}\sim\text{max}\left ( ||[H,O_{i_S}]||_2\right )$, with the jump operator $O_{i_S}$ defined below. 
Sampling from $\rho_E(\beta)$ is straightforward since the partition function factorizes.
Starting from $|0\rangle^{\otimes N_E}$, this is done by applying $X_{i_E}$ with probability,
\begin{equation}
   \text{Pr}(X_{i_E}) \ = \ \frac{1}{1+e^{\beta \omega_{i_E}}} \ .
   \label{eq:pBeta_E}
\end{equation}

Next, the system and environment are evolved for time $T$ with,
\begin{align}
    &U(T) \ = \ {\cal T}e^{-i \int_0^TdtH(t-T/2)}\ , \nonumber \\
    &H(t) \ = \ H_S \ + \ H_E \  + \ \alpha\, f(t)H_{SE}\ ,
    \label{eq:fullH}
\end{align}
where ${\cal T}$ is the time-ordering operator.
The total Hamiltonian $H(t)$ includes a time-dependent system-environment coupling term $H_{SE}$, with strength $\alpha$. 
The system-environment coupling has the form,
\begin{align}
H_{SE} \ = \  \sum_{\langle i_S,i_E \rangle} O_{i_S} \otimes X_{i_E}
\end{align}
where each environment qubit is coupled to a specific system qubit as denoted by $\langle i_S,i_E \rangle$. 
If $N_E<N_S$, then the pairing between environment and system qubits is chosen randomly at each reset cycle.
The jump operators $O_{i_S}$ induce transitions between the different eigenstates of $H_S$ and are normalized to $||O_{i_S}||_2 =1$.
Unless otherwise specified, the jump operators are chosen randomly from $O_{i_S}\in \{ X_{i_S},Y_{i_S}, Z_{i_S} \}$ at the beginning of each reset cycle. 
The filter function $f(t)$ is a Gaussian
\begin{align}
f(t) \  =\  {\cal N}\exp\left(-t^2/(4\sigma^2T^2)\right) \ ,
\label{eq:f(t)}
\end{align}
where ${\cal N}$ is the normalization factor.
The width of the Gaussian sets the energy resolution for inducing transitions between eigenstates and, for constant steady-state error, should scale with $\beta$.
Evolution under a time-independent Hamiltonian—corresponding to $\sigma=\infty$—is also valid, but results in worse scaling of the fixed-point error.
This is shown in  Appendix~\ref{app:csim}.

\begin{algorithm}[t]
\caption{Gibbs Sampling via Reset Channel}
\label{alg:algobox}
\KwIn{System Hamiltonian $H_S$, inverse temperature $\beta$, number of resets $N_{\mathrm{resets}}$}
Initialize system to a state $|\psi_{\text{in}}\rangle$\\
\For{$r=1$ \KwTo $N_{\mathrm{resets}}$}{
    Sample Bohr frequencies $\omega_{i_E}$ and jump operators $O_{i_S}$\\
    Prepare environment qubits in a product state sampled from $\rho_E(\beta)$\\
    Evolve system and environment under $H(t)=H_S+H_E+\alpha f(t) H_{SE}$ for time $T$\\
    Measure environment qubits\\
}
\KwOut{State on the system sampled from an approximate Gibbs ensemble}
\end{algorithm}

In practice, the time-dependent evolution is broken into steps of size $\delta_t$ and performed discretely e.g.,
\begin{align}
U(T) \ = \ \prod_{j=0}^{T/\delta_t - 1} e^{-i \delta_t H\left (-T/2 + (j+1/2)\delta_t \right )} \ .
\label{eq:UT_Trotter}
\end{align}
Each term in the product is digitized into gates using, for example, a second-order Trotterization.
The normalization ${\cal N}$ is fixed by \cite{chen2023quantum},
\begin{align}
\sum_{j=0}^{T/\delta_t - 1} \delta_t f(-T/2 + (j+1/2)\delta_t)^2 \ = \ 1 \ .
\end{align}
This ensures that the strength of the dissipative channel is independent of $T$.
After evolving under $U(T)$, the environment qubits are measured and reset to $|0\rangle^{\otimes N_E}$.
This process is repeated $N_{\text{resets}}$ times, and the pure states produced $\{|\psi_{\text{out}}\rangle\}$  are sampled from an approximate Gibbs ensemble.

Each reset cycle implements a quantum channel $\Phi$ that acts on the system qubits,
\begin{equation}
    \Phi(\rho_S) =\Tr_E[U(T)(\rho_S \otimes \rho_E) U^\dagger(T)] \ . 
    \label{eq:qChannel}
\end{equation}
Implicit in this equation is an average over the randomized $\omega_{i_E}$ and $O_{i_S}$.
It was shown in Ref.~\cite{Ding:2025ulc} that the unique fixed point of this channel, $\tilde{\rho}_S(\beta)$, approximates the target Gibbs state with an error $\|\tilde{\rho}_{S}(\beta)-\rho_{S}(\beta)\|_1<\epsilon$.
This was proved by analyzing the steady state of the quantum channel in Eq.~\eqref{eq:qChannel} expanded to ${\cal O}(\alpha^2)$,
\begin{align}
\Phi(\rho_S) \ = \ \rho_S \ + \ \alpha^2\mathcal{L}(\rho_S) \ + \ \mathcal{O}(\alpha^4) \ .
\label{eq:PhiExpand}
\end{align}
The second-order term $\mathcal{L}(\rho_S)$ is a Lindbladian that satisfies an approximate quantum detailed balance condition, under which the thermal state is an approximate fixed point \cite{chen2025efficient, Ding:2025ulc}.
Recent work further shows that the higher order terms also satisfy the approximate detailed balance condition \cite{wang2025beyond}. 

Classical detailed balance posits that the  transition rates between energy eigenstates $|E_i\rangle$ and $|E_j\rangle$ satisfy
\[
\frac{\operatorname{rate}(i \rightarrow j)}{\operatorname{rate}(j \rightarrow i)}=e^{-\beta\left(E_j-E_i\right)} \ .
\]
Quantum detailed balance generalizes this condition to coherent transitions between energy eigenstates.
The quantum detailed balance condition is not unique and one choice is that
the action of a Lindbladian acting on a density matrix ${\cal L}(\rho) = d\rho/dt$ satisfies~\cite{chen2025efficient}
\begin{align}
&\left\langle E_i\right| \mathcal{L}\left( \left|E_j \right\rangle\left\langle E_k \right|\right) \left|E_l \right\rangle= \nonumber  \\ & \exp \left(-\beta \frac{E_k-E_l+E_j-E_i}{2}\right)\left(\left\langle E_i\right| \mathcal{L}\left( \left|E_j \right\rangle\left\langle E_k \right|\right) \left|E_l \right\rangle\right)^* \ .
\label{eq:QuantumDetailedBalance}
\end{align}
This condition constrains the transitions induced by the Lindbladian between both the diagonal and off-diagonal elements of the density matrix in the energy basis.
It reduces to the classical detailed balance condition if the density matrix is diagonal in the energy basis ($\left|E_j\right\rangle=\left|E_k\right\rangle$) and the Lindbladian only induces transitions between energy eigenstates ($\left|E_i \right\rangle= | E_l \rangle$). 
The fixed-point error of the prepared state $\epsilon$ depends on how well the quantum channel is reproduced by digital gates through $\{\delta_t,T\}$, as well as the deviation from the detailed balance condition through $\{\alpha,\sigma,T\}$.
The number of channel applications $N_{\mathrm{resets}}$ required to reach the steady state is governed by the mixing time $\tau_{\mathrm{mix}}$, which depends on the system Hamiltonian $H_S$, the inverse temperature $\beta$, and the interaction strength $\alpha$. 

To ensure the uniqueness of the fixed point, the jump operators are chosen such that the only operator that commutes with all of the $O_{i_S}$ is the identity. The simplest choice is the set of all single-qubit Pauli operators. 
Our numerical results in Section~\ref{sec:classicalsims} show that this is sufficient for Gibbs sampling in the AFIM. 
However, when the Hamiltonian possesses internal symmetries, the jump operators must be chosen carefully to either break or preserve the symmetries. 
For example, in fermionic systems, the jump operators $\{a_i^\dagger a_j\}_{i,j=1}^{N_S}$ are required to prepare the Gibbs state in the canonical ensemble, which preserves particle number, while $\{a_i^\dagger, a_i\}_{i=1}^{N_S}$ are needed for the grand canonical ensemble \cite{li2025dissipative}. 
In our AFHM results, we show that additional $SU(2)$ symmetry-preserving jump operators are needed to reach thermal equilibrium at low temperatures.
This is discussed further in Methods~\ref{methods:jump}.

Successful implementation of this algorithm on current, noisy, quantum computers requires a careful balancing of the fixed point error $\epsilon$ against the required quantum resources.
This amounts to a tuning of $\{\alpha,T,\sigma,\delta_t\}$ to maximize the quality of the prepared state in the presence of device noise. Although the parameter choices to achieve polynomial mixing time (when possible) is given in Ref.~\cite{Ding:2025ulc}, the optimal parameter tuning in practice is heuristic, and can be informed from classical statevector simulations on small lattices.

\subsection{Low-temperature properties of the AFIM and AFHM}
\label{methods:FrustratedPhysics}
\noindent
The AFIM on the kagome lattice is a paradigmatic model in the field of frustrated magnetism, whose key physics is understood following the seminal work by Moessner {\it et al}~\cite{Moessner_2000, Moessner_2001, Moessner2001_dimer} and which we follow. 
If only a longitudinal field is present (the classical model), energy constraints per triangle require a 2-up 1-down structure on every triangle (or vice versa if the longitudinal field points in the opposite direction). 
Such a local constraint is akin to a Gauss law in gauge theories, and is called the ice rule. It can be satisfied by exponentially many configurations, giving rise to a finite entropy density. 
This phase is a classical spin liquid known as kagome ice. 
It can be identified through characteristic pinch points in the static structure factor, reminiscent of a Coulomb phase.
In the absence of a longitudinal field, the system remains a quantum paramagnet (or trivial ferromagnet in the transverse direction) down to zero temperature, irrespective of the value of the transverse field. 
This is due to a mechanism known as disorder-by-disorder. 
A quantum paramagnet is not a quantum spin liquid; it has short-range entanglement, trivial excitations and no topological order. 
In fact, pure $U(1)$ quantum spin liquids in two dimensions are known to be unstable to perturbations~\cite{Polyakov1977}.
In the presence of weak longitudinal and transverse fields, the system can remain in a renormalized classical kagome spin ice down to extremely low temperatures. 
Through a mapping onto a quantum dimer model, the model is known to be close to the Rokhsar-Kivelson quantum spin liquid~\cite{RK1988}, yet expected to form a valence bond solid in the ground state. Indeed, through a mechanism known as order-by-disorder at a scale set by sixth order degenerate perturbation theory around a hexagon, a valence-bond solid with a maximal number of flippable hexagons and a $\sqrt{3} \times \sqrt{3}$ unit cell is predicted.

By contrast, the physics of the AFHM on the kagome lattice remains intensely debated. 
We cannot and do not intend to give a full overview; rather, we just select some of the main numerical developments. 
Path integral Monte Carlo simulations suffer from the infamous sign problem and cannot be meaningfully applied.
High temperature series expansions have been carried out to orders 17-20 for the specific heat and spin susceptibility~\cite{BernuLhuillier2015, Bernu2020}. They converge for temperatures $\beta\lesssim 1$ 
with Pad{\'e} extrapolation giving consistent results up to $\beta\lesssim 2.5$.
The specific heat consistently resolves a bump in a high-temperature region around 
$\beta\sim 1$;
depending on the nature of the ground state additional features might exist in a second, low-temperature crossover region for $\beta \gtrsim 10$. 
DMRG studies from around 2011 claimed a gapped $Z_2$ quantum spin liquid ground state~\cite{Yan:2010tty,Depenbrock2012} with variational energies that are much lower than the ones of a valence bond crystal~\cite{Singh2007,Evenbly2010}.  
However, contemporaneous variational Monte Carlo methods found a gapless Dirac spin liquid~\cite{Ran2005, Iqbal2011, Iqbal2013}. The iDMRG calculations from 2017 showed through flux insertion techniques that the spin gap is much smaller than the ones extracted from the earlier DMRG studies.
In fact, the iDMRG calculations also point to a $U(1)$ Dirac spin liquid~\cite{He2017}, which can be seen as a parent state to many of the other states, including chiral spin liquids. The Dirac spin liquid is currently the most widely accepted scenario.

\subsection{Jump operator connectivity}
\label{methods:jump}
\noindent
Convergence of the dissipative quantum channel to a thermal ensemble requires that every pair of energy eigenstates is connected by a sequence of jump operators. 
Specifically, for every pair of eigenstates $|E_n\rangle,|E_m\rangle$
\begin{align}
\langle E_n| O_i O_j O_k \ldots |E_m\rangle
\end{align}
is nonzero for some $\{i,j,k,\ldots\}$.
Effective jump operators arising at higher orders in the system-bath coupling that are generated through time evolution are neglected in this expression.
Thermalization is fast when there are many jump sequences with large amplitudes that connect the energy eigenstates.
In many systems, the bottleneck is at low energies where the density of states is the smallest.
This changes in frustrated systems because there are exponentially many low-energy states.
Instead, the bottlenecks are the connections between the manifold of nearly-degenerate low-energy states and the rest of the spectrum.

\begin{figure}
    \centering
    \includegraphics[width=\linewidth]{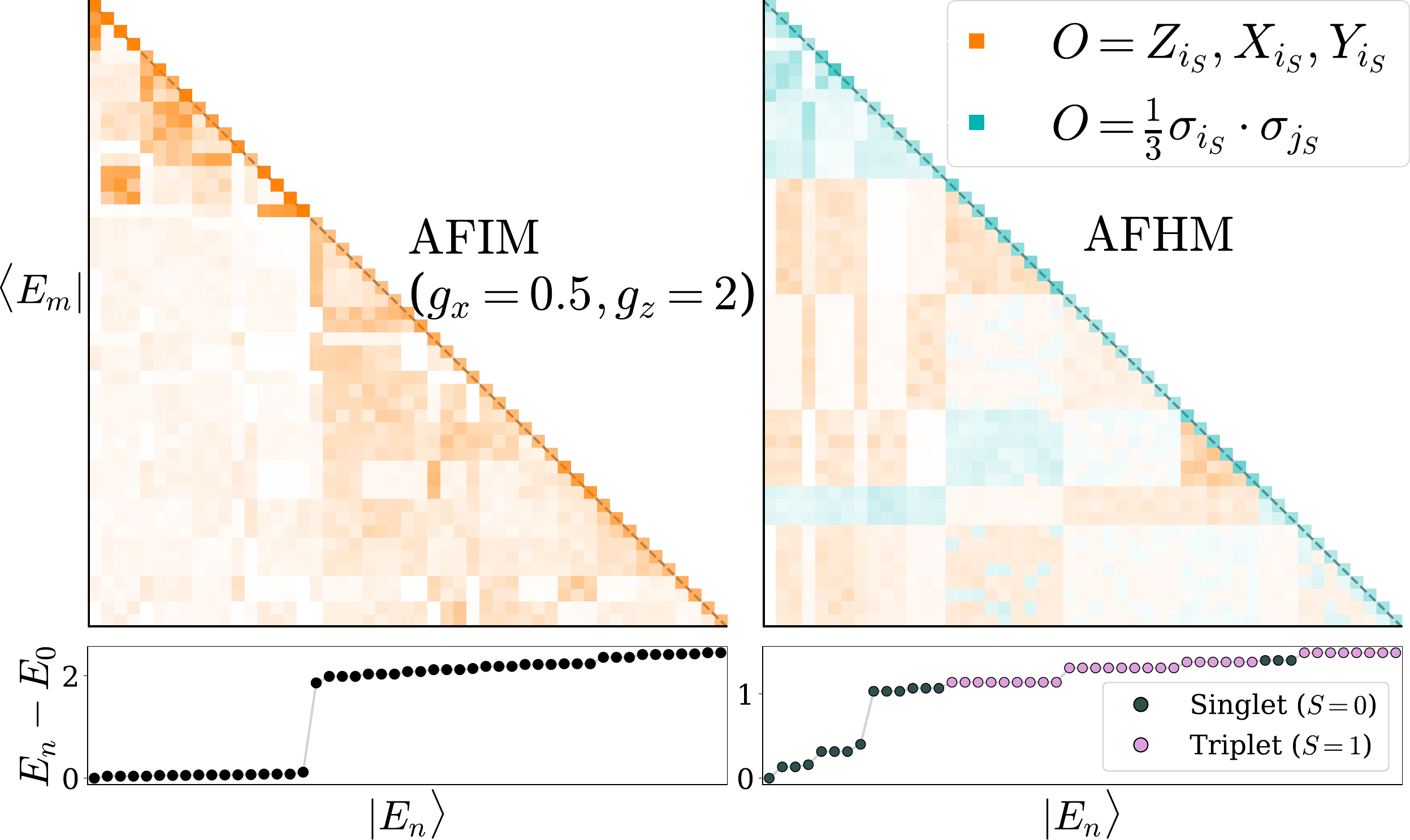}
    \caption{
    Matrix elements of the jump operators $O$ between the 50 lowest-lying energy eigenstates of the AFIM and AFHM on a $N_S=18$ site kagome lattice with PBCs. 
    The opacity of the pixel at $(n,m)$ represents $\text{max}\left  (|\langle E_m|O|E_n\rangle|\right )$ where the maximum is taken over all jump operators.
    The pixels are orange if the maximum is a single site jump operator and teal if it is a $SU(2)$ singlet jump operator.
    The opacity has been normalized by the largest entry for each of the two models independently.
    The bottom panels show the energy of the corresponding eigenstates.
    }
    \label{fig:JumpConnectivity}
\end{figure}

The jump operator connectivity matrix between a selection of low energy eigenstates is shown in Fig.~\ref{fig:JumpConnectivity} for the AFIM (left) and AFHM (right).
The connectivity matrix of $O_{i_S}\in \{X_{i_S},Y_{i_S},Z_{i_S}\}$ in the AFIM is approximately block diagonal between the ice and non-ice subspaces.
Within the ice manifold, eigenstates are well-connected by local operators (primarily $Z_{i_S}$) and  the same is true within the non-ice subspace. 
However, the connection between ice and non-ice eigenstates is weak. 
This has important consequences for the mixing time as discussed below.

The connectivity matrix in the AFHM is additionally constrained by the global $SU(2)$ symmetry.
The $O_{i_S}\in \{X_{i_S},Y_{i_S},Z_{i_S}\}$ operators have spin $S=1$ and therefore cannot connect $S=0$ states in a single jump.
Matrix elements between other irreps are generally non-zero as ${\bm S}\otimes{\bm 1} = {\bm S}\oplus\ldots$ for ${\bm S}\neq {\bm 0}$.
As a result, the $S=0$ subspaces are not connected by a single application of $O_{i_S}$.
This slows down thermalization for large $\beta$ as transitions between the low-energy, singlet, configurations require multiple jumps. 
To address this problem we also include $S=0$ jump operators
\begin{align}
O_{i_S,j_S}=\frac{1}{3}\left ( X_{i_S}X_{j_S}+Y_{i_S}Y_{j_S}+Z_{i_S}Z_{j_S} \right ) \ ,
\label{eq:Osinglet}
\end{align}
for all pairs of sites $i_S, j_S$. Non-local jumps with $i_S,j_S$ beyond nearest neighbor are needed to connect the low-energy eigenstates with a single jump.
These jump operators connect the $S=0$ states as illustrated in Fig.~\ref{fig:JumpConnectivity}.

In Section~\ref{sec:classicalsims}, it was shown that the mixing time in the AFIM  has a local maxima at the crossover temperature $\beta_c\approx 1.4$.
The corresponding energy $E= 1/\beta_c \approx 0.7$ is in the middle of the energy gap separating the ice and non-ice states, see the bottom left panel of Fig.~\ref{fig:JumpConnectivity}.
The slow mixing at $\beta_c$ is likely due to the jump operator connectivity being approximately block diagonal between the ice and non-ice sectors.
At the fixed point, the quantum detailed balance condition in Eq.~\eqref{eq:QuantumDetailedBalance} requires the Lindbladian to maintain specific transition rates between all thermally occupied eigenstates. 
For $\beta>\beta_c$, only ice states have significant thermal occupation, and detailed balance only requires transitions within the ice manifold where the jump operators have large matrix elements. 
At $\beta\approx \beta_c$, both ice and non-ice states are thermally occupied in comparable amounts, and detailed balance demands transitions between the two sectors. 
The weak matrix elements connecting ice and non-ice states suppress these transitions creating a bottleneck that increases the mixing time.
The scaling of the classical computing resources required for these simulations prevented a study of how this slowdown scales with system size.
Investigating the sharpness of this slowdown for increasing system sizes is an exciting application for future quantum computers.

\subsection{Setup for thermal states prepared on IBM's quantum computers}
\label{methods:qsim}
\noindent
In this section, we provide the design choices made in the implementation of the thermal state preparation algorithm on {\tt ibm\_boston}.
The dissipative algorithm for sampling from Gibbs states is described in Methods~\ref{methods:algo} and illustrated in Fig.~\ref{fig:overview}. 
First, the system qubits are initialized to a state drawn from $\{|\psi_{\text{in}}\rangle\}$.
$\{|\psi_{\text{in}}\rangle\}$ is chosen to be a random  product state ($\beta=0$) in the computational basis.
Each $|\psi_{\text{in}}\rangle$ is prepared by applying $X_{i_S}$ to each system qubit $i_S$ with probability $\text{Pr}(X_{i_S})=1/2$.
Second, the Bohr frequencies of the environment Hamiltonian are randomly sampled from $\omega_{i_E}\in (0,\omega_{\text{max}}]$.
We set $\omega_{\text{max}}=\text{max}(4g_x, 4g_z,4) = 8$ corresponding to twice the bandwidth of the individual terms in the Hamiltonian~\cite{Lloyd:2025cvp}.
The rate of convergence to the steady state is insensitive to small variations of $\omega_{\text{max}}$.
The environment qubits are prepared in a product state sampled from $\rho_E(\beta)$ in Eq.~\eqref{eq:HE}.
This is done by applying $X_{i_E}$ gates to each environment qubit $i_E$ with probability $\text{Pr}(X_{i_E})$ that is given in Eq.~\eqref{eq:pBeta_E}.

Next, jump operators are randomly sampled from $O_{i_S}\in\{ X_{i_S}, Y_{i_S},  Z_{i_S} \}$ and the joint system-environment is evolved under the $U(T)$ in Eq.~\eqref{eq:UT_Trotter}.
We choose $T=0.75$ and $\delta_t=0.25$ to balance quantum resources and fixed point error, requiring $T/\delta_t=3$ Trotter steps per reset cycle.
These choices, along with a system-environment coupling of $\alpha =1.75$, are informed from classical statevector simulations on a $N_S=12$ kagome lattice as discussed in Appendix~\ref{app:csim}.
With this few Trotter steps, a constant filter function ($\sigma=\infty$) is optimal.
Thus, $U(T)$ is time independent and we implement it with a second order Trotterization,
\begin{align}
U(T) \ = \ \prod_{j=1}^{T/\delta_t } &e^{-i H_{SE} \delta_t /2}e^{-i H_{S,X} \delta_t /2} e^{-i H_{S,Z} \delta_t }e^{-i H_{S,ZZ} \delta_t}\nonumber \\
&\times e^{-i H_{E} \delta_t}e^{-i H_{S,X} \delta_t /2}  e^{-i H_{SE} \delta_t /2} \ .
\end{align}
The system Hamiltonian in Eq.~\eqref{Eq:HKagome} has been split into three pieces $H_S = H_{S,Z}+H_{S,ZZ}+H_{S,X}$, representing the terms with single qubit $Z$, $X$ and two qubit $ZZ$.
The single-qubit terms, including those in $H_E$, are implemented with single-qubit rotations.
Evolution under $H_{SE}$ includes two-qubit terms of the form $\sigma_{i_S}X_{i_E}$ that couple each environment qubit to its corresponding system qubit.
The environment and system qubits are natively coupled on {\tt ibm\_boston} using the layout in Fig.~\ref{fig:ibm_circuits}{\bf a}, and evolution under each term in $H_{SE}$ is implemented with single-qubit rotations and a single $R_{ZZ}(\theta)$ two-qubit gate.
The combined evolution under the single-qubit terms and $H_{SE}$ is represented by the green chain-link in Fig.~\ref{fig:ibm_circuits}{\bf b}.

\begin{figure}
    \centering
    \includegraphics[width=\linewidth]{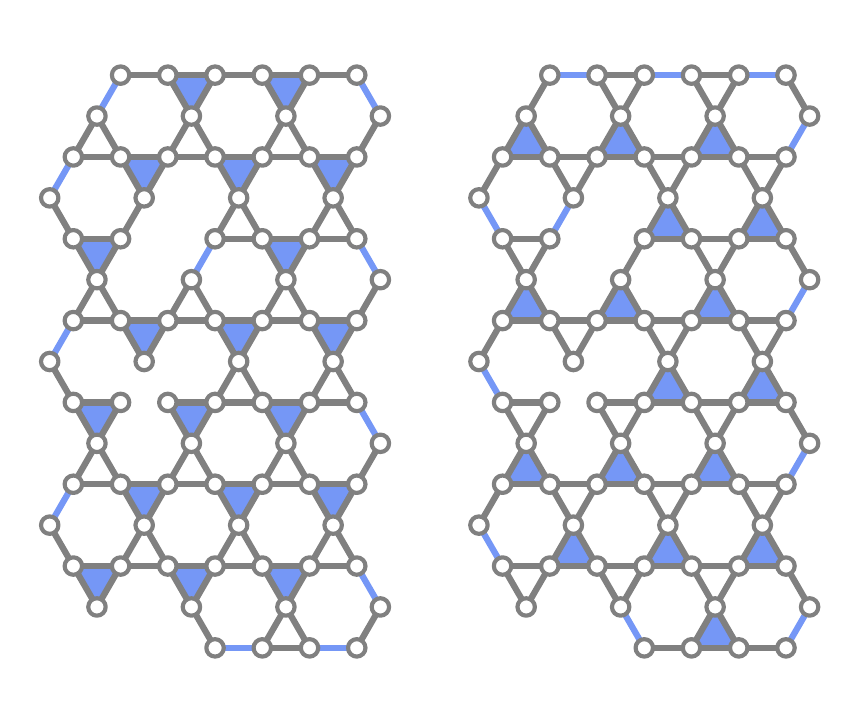}
    \caption{The partitioning of the links on the $N_S=79$ kagome lattice into two sets for implementing Trotterized time evolution.
    The left (right) set contains all downward (upwards) pointing triangles. 
    Boundary links are added to a set if they do not share an endpoint with one of the triangles.}
    \label{fig:kagome_coloring}
\end{figure}
More difficult is evolution under $H_{S,ZZ}$, which couples all qubits connected by the pink links of the kagome lattice in Fig.~\ref{fig:ibm_circuits}{\bf a}.
These sites are not natively coupled on heavy-hex and interactions have to be routed through the environment qubits.
The best circuit structure we have found for reducing both the number of two-qubit gates and the gate depth is illustrated in Fig.~\ref{fig:ibm_circuits}{\bf b} and {\bf c}.
First, the triangles that make up the kagome lattice are partitioned into those that point up and those that point down.
The links on the boundaries are added to the set of the triangles that they do not share endpoints with.
This partitioning of links is shown for the $N_S=79$ kagome lattice used on {\tt ibm\_boston} in Fig.~\ref{fig:kagome_coloring}.
The blue triangles implements $R_{ZZ}(\theta)$ between the qubits on its vertices and the blue links implements $R_{ZZ}(\theta)$ between the qubits on its endpoints.
On heavy-hex, these gates have to be routed through an environment qubit, and the gate sequence used is shown in Fig.~\ref{fig:ibm_circuits}{\bf c}.
The triangle circuit first uses a swap gate between the environment and a system qubit so that the three vertices are connected in a line.
It then uses the nested $R_{ZZ}(\theta)$ circuit from Ref.~\cite{Farrell:2024fit} to implement $R_{ZZ}(\theta)$ between all pairs of qubits.
Lastly, another swap returns the qubits to their original orientation.
The system qubit that is swapped is arbitrary, and therefore the white vertices of each triangle in Fig.~\ref{fig:ibm_circuits}{\bf c} can be permuted without changing the unitary.
This permutation freedom is used to reduce the two-qubit gate depth as explained in Appendix~\ref{app:qcircs}.
Additional quantum circuits for implementing Trotterized evolution in the AFHM are also provided in Appendix~\ref{app:qcircs}.

After evolving under $U(T)$, the environment qubits are measured and conditionally restored to $|0\rangle$ using {\tt reset} instructions.
This quantum channel is repeated for $N_{\text{resets}}$ cycles. 
At the start of each cycle, random Bohr frequencies $\omega_{i_E}$ and jump operators $O_{i_S}$ are drawn, and the environment qubits are initialized to a fresh sample drawn from $\rho_E(\beta)$.
The number of randomizations per point is 250 for $N_\mathrm{resets}=1$, where the observables exhibit significant variance due to the random initialization of the system qubits at the beginning of each shot. 
As few as 50 randomizations were used at $N_\mathrm{resets}=7$, where the fluctuations are smaller. 
Each randomization has 200 shots, of which as few as $22\%$ survive leakage post-selection.
The error mitigation strategies are detailed in Appendix~\ref{app:errormitigation}.

\onecolumngrid

\clearpage
\appendix
\section{Rate of convergence to the steady state with varying number of environment qubits}
\label{app:nQ_convergence}
\begin{figure}
    \centering
    \includegraphics[width=\linewidth]{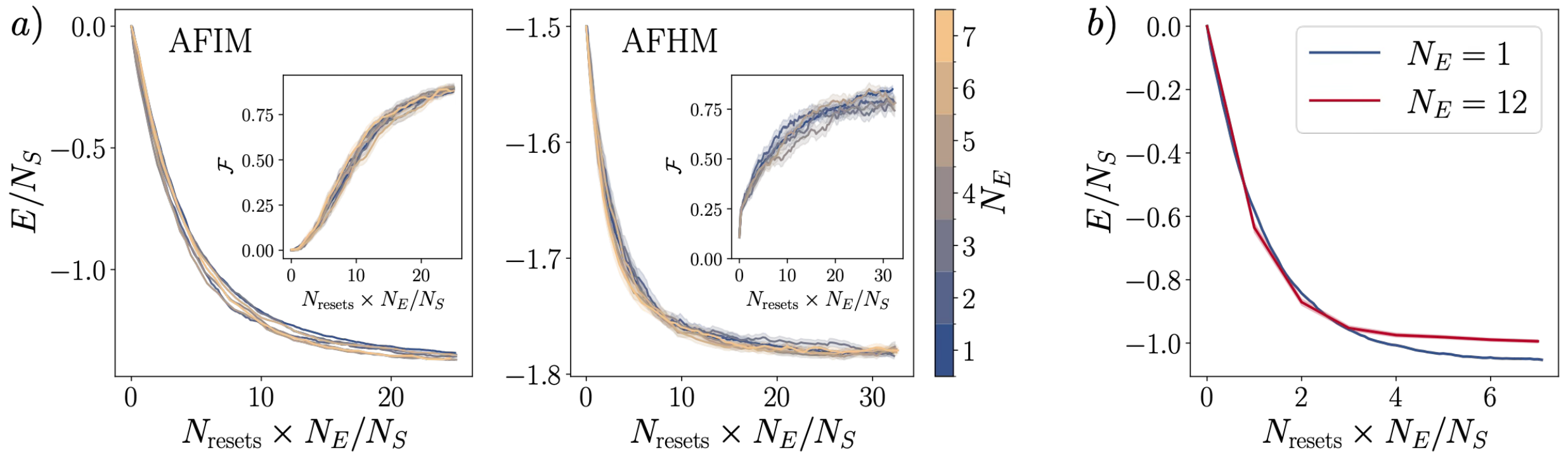}
    \caption{The energy density as a function of the number of resets $N_{\text{resets}}$ on a $N_S=12$ kagome lattice with varying number of environment qubits $N_E$.
    a) The left plot is for $\beta=3$ in the AFIM and the right plot is for $\beta = 4$ in the AFHM, both with PBCs.
    The AFIM starts from $\beta=0$ whereas the AFHM starts from a tensor product of $SU(2)$ singlets.
    The inset shows the corresponding fidelity ${\cal F}$ with the exact thermal state.
    The fidelity in the AFHM is only computed for $N_E<6$ due to classical computing limitations.
    All trajectories have 200 samples except for $N_E=1$ which has 1000 (750) trajectories in the AFIM (AFHM).
    b) The energy density for $N_E=1$ and $N_E=12$ using the same parameters as in the experiments run on {\tt ibm\_boston}.
}
    \label{fig:nBathscaling}
\end{figure}

\noindent
The number of environment qubits in the dissipative algorithm can be tuned over the range $N_E\in[1,2,\ldots,N_S-1,N_S]$.
Increasing $N_E$ enhances the number of couplings between the system and environment and therefore reduces the number of resets required for thermalization, i.e., the mixing time.
We verify this with statevector simulations on an $N_S=12$ kagome lattice with PBCs.
The parameters chosen are the same as those in Section~\ref{sec:classicalsims} and are given in the caption of Fig.~\ref{fig:IsingHeisFid}.
The initial states in the AFIM are sampled from random $(\beta=0)$ product states.
To reduce the number of resets required to reach the steady state in the AFHM, we set the initial state to be a tensor product of nearest-neighbor $SU(2)$ singlet dimers, as explained in Section~\ref{sec:classicalsims}.
The energy density as a function of the number of resets is shown in Fig.~\ref{fig:nBathscaling}a) for $\beta=3$ in the AFIM and $\beta=4$ in the AFHM.
The fidelity with the exact thermal state is shown in the inset.
Rescaling the number of resets by $N_E/N_S$ collapses all trajectories onto a single curve, removing the dependence on $N_E$.
The collapse reflects two features of the dissipative dynamics:
the mixing time is $\tau_{\text{mix}}\propto1/N_E$ and the steady state is approximately independent of $N_E$.
These features hold when each application of the quantum channel is well described by the leading-order-in-$\alpha$ thermalizing Lindbladian in Eq.~\eqref{eq:PhiExpand}.

Subleading terms in the expansion of the quantum channel in Eq.~\eqref{eq:PhiExpand} become non-negligible for large $\alpha$ and cause the steady state to depend on $N_E$.
This effect is non-negligible in the experiments run on IBM's quantum computer that have $\alpha=1.75$.
This is illustrated in Fig.~\ref{fig:nBathscaling}b) which gives the energy density for $N_S=12$ and both $N_E=1$ and $N_E=12$.
While the convergence rate is roughly independent of $N_E$, the steady-state energy at $N_E=1$ is approximately $5\%$ lower than at $N_E=12$.
Accurate comparison with experiment therefore requires noiseless simulations performed at the same $N_E$.
Accordingly, the noiseless curves in Fig.~\ref{fig:IBM_results_Energy_nE} use $N_E=12$ for $N_S=12$, matching the experimental setup.
However, due to classical memory constraints, the experimental results for $(N_S,N_E)=(18,17)$ are compared to statevector simulations with $(N_S,N_E)=(18,14)$.
Consequently, the reported steady-state energy is likely slightly lower than the true noiseless value.
%

\section{Mixed state fidelity}
\label{app:fid}
\begin{figure}
    \centering
    \includegraphics[width=0.75\linewidth]{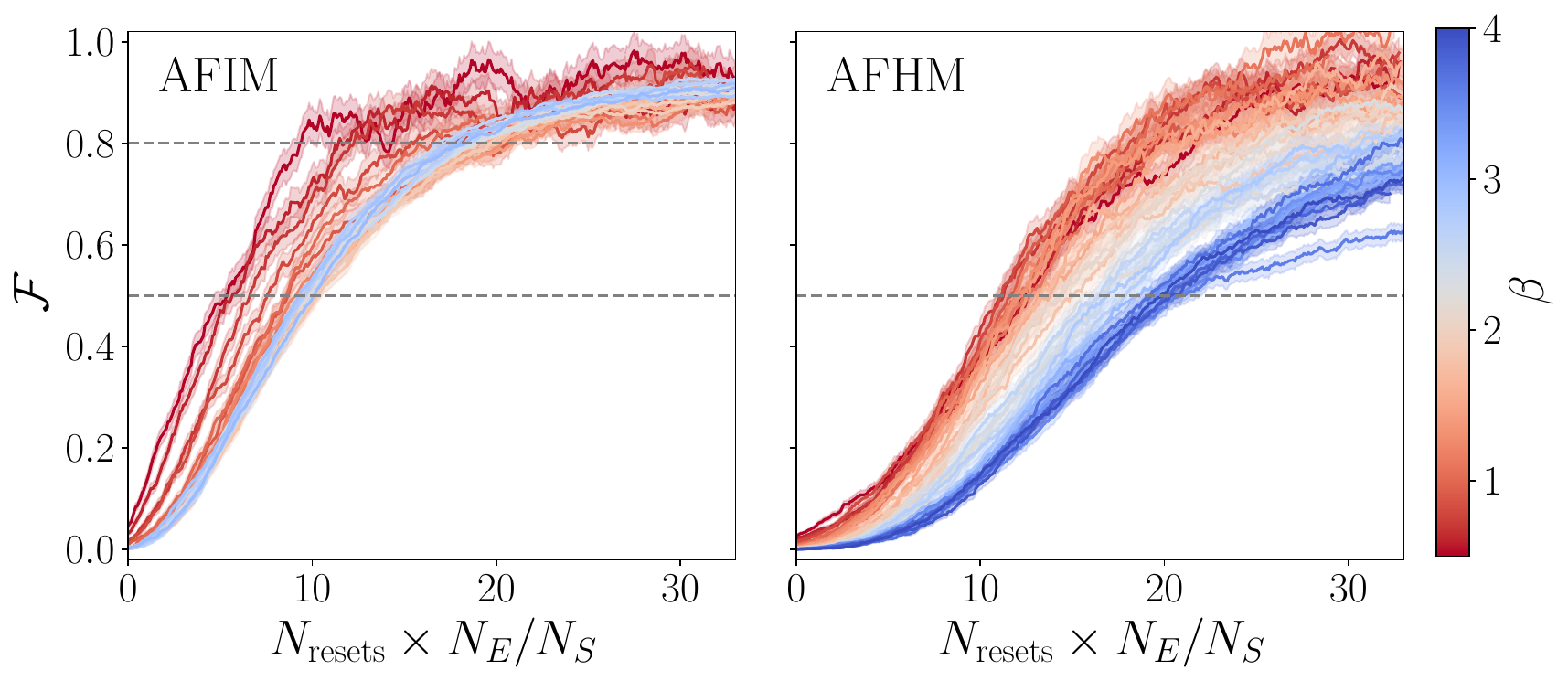}
    \caption{The mixed state fidelity ${\cal F}$ between the prepared state and the exact thermal density matrix on a $N_S=12$ kagome lattice with PBCs.
    Left: Results from 1000 samples in the AFIM with $T=8,\sigma^2=1/16,\omega_{\text{max}}=8$.
    Right: Results from 750 samples in the AFHM with $T=14,\sigma^2=1/24,\omega_{\text{max}}=4$.
    Both plots have $\delta_t=0.1,\alpha=1,N_E=1$ and sample $\{|\psi_{\text{in}}\}$ from a random ($\beta=0$) product state.
    In Fig.~\ref{fig:mix_scaling}, the AFIM and AFHM sample (1000, 200, 40) and (750, 200, 40) trajectories respectively for $N_S=(12,18,24)$.
    For $N_S=27$, 60 (20) trajectories were sampled for the AFIM (AFHM).}
    \label{fig:IsingHeisFid}
\end{figure}

\noindent
Figure~\ref{fig:mix_scaling}{\bf b} gives the mixing time, defined as the number of resets needed to reach a threshold fidelity ${\cal F}_{\text{thresh}}$. 
To provide more detail on the rate of convergence, we additionally show the fidelity for increasing number of reset steps in Fig.~\ref{fig:IsingHeisFid}.
The dotted lines correspond to the two values of ${\cal F}_{\text{thresh}}$ used, and their intersection with the fidelity curves correspond to the mixing time $\tau_{\text{mix}}({\cal F}_{\text{thresh}})$.
The crossing of the light-blue and light-red curves in the AFIM gives rise to the local maxima in the mixing time at $\beta_c\approx1.4$.

The fidelity between the prepared state $\tilde{\rho}_S(\beta)$ and the exact Gibbs state $\rho_S(\beta)$ throughout this work is computed with the mixed-state fidelity metric,~\cite{Liang:2018yey}
\begin{align}
{\cal F}(\rho_1, \rho_2) \ = \ \frac{\text{Tr}(\rho_1 \, \rho_2)}{\text{max}\left [\text{Tr}(\rho_1^2),\, \text{Tr}(\rho_2^2) \right ]} \ . 
\label{eq:mixedFid}
\end{align}
Constructing the full $2^{N_S}\times 2^{N_S}$ is infeasible beyond $N_S=12$ and we instead use lower-rank approximations.
For $\rho_S(\beta)$ we use,
\begin{align}
\rho_S(\beta) \ &\approx \ \left (\sum_{E_i-E_0<c/\beta}  e^{-\beta E_i}|E_i\rangle\langle E_i|\right ) \bigg /\left (\sum_{E_i-E_0<c/\beta}e^{-\beta E_i}\right )  \nonumber \\[4pt]
&= \ \frac{1}{Z_I}\sum_{i=0}^{I-1}  e^{-\beta E_i}|E_i\rangle\langle E_i| 
\end{align}
where $I$ is the number of energy eigenstates contributing to the approximate density matrix and $Z_I=\sum_{i=0}^{I-1}e^{-\beta E_i}$ is the approximate partition function.
The approximation error is controlled by the truncation parameter $c$.
Higher energy eigenstates are exponentially suppressed by $e^{-\beta (E_i-E_0)}$ but have a higher density of states.
This truncation would break down for low temperatures that probe the quantum spin liquid phase as the density of states grows exponentially over the frustrated manifold.
However, for the modest sized lattices we explore, we find that the cutoff parameter $c=8$ is well converged.
For $\tilde{\rho}_S(\beta)$, the density matrix is reconstructed from the sampled pure states,
\begin{align}
    \tilde{\rho}_S(\beta) \ \approx \ \frac{1}{J}\sum_{j=0}^{J-1} |\psi_{\text{out}}^{(j)}\rangle\langle\psi_{\text{out}}^{(j)}| \ . 
\end{align}
The state $|\psi_{\text{out}}^{(j)}\rangle$ is the output of a single trajectory of the quantum channel.

With these low-rank approximations, the purity can be estimated without explicitly forming the $2^{N_S}\times 2^{N_S}$ density matrices.
Inserting the approximations into Eq.~\eqref{eq:mixedFid} gives,
\begin{align}
{\cal F}\left [\tilde{\rho}_S(\beta), \rho_S(\beta)\right ] \ \approx \ \frac{\frac{1}{J\, Z_I}\sum_{i=0}^{I-1}\sum_{j=0}^{J-1}e^{-\beta E_i}|\langle E_i|\psi_{\text{out}}^{(j)}\rangle |^2 }{\text{max} \left [ \left ( \sum_{i=0}^{I-1}e^{-2\beta E_i}\right )/Z_I^2\, ,  \text{Tr}\left (\tilde{\rho}^2_S(\beta)\right )\right]}
\label{eq:approxFid}
\end{align}
where, 
\begin{align}
\text{Tr}\left (\tilde{\rho}^2_S(\beta)\right ) \ = \ \left ( \sum_{j=0}^{J-1}\sum_{j'=0}^{J-1}\left ( |\langle \psi_{\text{out}}^{(j)}|\psi_{\text{out}}^{(j')}\rangle |^2\right ) - J\right )/(J^2-J) \ ,
\end{align}
These expressions only require evaluating overlaps between $2^{N_S}$-dimensional vectors and can be vectorized for efficient computation.
The offset in the purity of $\tilde{\rho}^2_S(\beta)$ accounts for the positive sampling bias that comes from the diagonal terms, which always contribute $1/J$, and would be present even if, e.g., $\tilde{\rho}_S(\beta) = \mathds{1}$.
In many of our classical simulations, the initial state is sampled from the infinite temperature Gibbs ensemble, which has vanishing purity.
In this case we observe that $  \text{Tr}\left (\tilde{\rho}^2_S(\beta)\right ) < \left ( \sum_{i=0}^{I-1}e^{-2\beta E_i}\right )/Z_I^2$ and the denominator in Eq.~\eqref{eq:approxFid} is a constant.
This is used to simplify many of our fidelity calculations.

The error bands on the fidelities are computed from the standard error of the mean of $\sum_{i=0}^{I-1}e^{-\beta E_i}|\langle E_i|\psi_{\text{out}}^{(j)}\rangle |^2$ across samples $j$, treating the denominator in Eq.~\eqref{eq:approxFid} as a constant.
The error bars on the mixing times in Fig.~\ref{fig:mix_scaling}{\bf b} is the range between the smallest $N_{\text{resets}}$ that satisfies
$({\cal F}(N_{\text{resets}})+1\sigma_{\text{sd}}) >{\cal F}_{\text{thresh}}$ and the largest $N_{\text{resets}}$ that satisfies ${\cal F}(N_{\text{resets}})<{\cal F}_{\text{thresh}}$ where $\sigma_{\text{sd}}$ is the standard deviation.

\section{Gate orientation to reduce the two-qubit gate depth of Trotterized time evolution}
\label{app:qcircs}
\begin{figure}[t]
\centering
\begin{minipage}{0.39\textwidth}
    \centering
    \includegraphics[width=\textwidth]{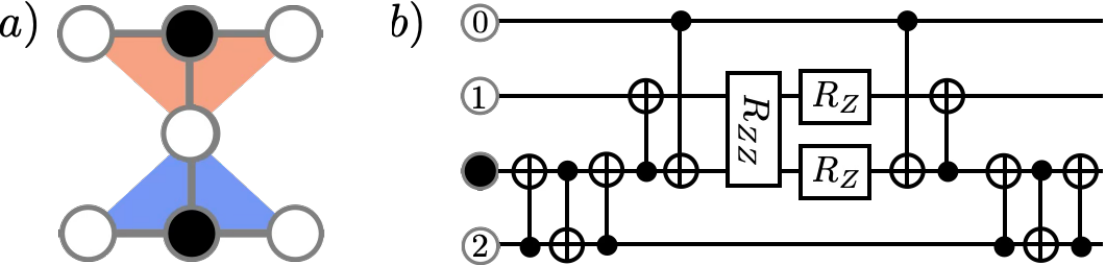}
\end{minipage}
\hfill
\begin{minipage}{0.6\textwidth}
    \centering
    \renewcommand{\arraystretch}{1.6}
    \begin{tabular}{|c||c|c|c|c|c|c|c|c|c|} \hline
    Shared triangle vertex&0 0&0 1&0 2&1 0&1 1&1 2&2 0&2 1&2 2 \\ \hline
    2q gate depth&12&15&17&15&15&18&17&18&22 \\ \hline
    \end{tabular}
\end{minipage}
\caption{Left: a) Trotterized time evolution under the AFIM on the kagome lattice requires $R_{ZZ}(\theta)$ between the vertices of each triangle.
The up (down) triangles have been colored blue (brown) and the time evolution circuit first performs the blue $R_{ZZ}(\theta)$s followed by the brown one.
b) The decomposition of the triangle using the native gate set and connectivity of IBM's {\tt heron} quantum computers. 
The vertices are labeled $0,1,2$ to track the orientation of each triangle.
Right: The upper row gives every combination of overlapping vertices between two triangles. 
The lower row gives the corresponding two-qubit gate depth.}
\label{fig:TriangleZZcirc}
\end{figure}

\noindent
The quantum simulations presented in Section~\ref{sec:qsims} used the circuit decomposition in Fig.~\ref{fig:ibm_circuits}{\bf b} and {\bf c} to implement the $ZZ$ terms in the AFIM Hamiltonian in Eq.~\eqref{Eq:HKagome}.
As explained in Methods~\ref{methods:qsim}, the kagome lattice is partitioned into up- and down-pointing triangles, and the $R_{ZZ}(\theta)$ gates are implemented between the qubits on the vertices of the triangles.
The vertices of the triangle can be permuted without changing the unitary and this freedom can be used to minimize the two qubit gate depth as illustrated in Fig.~\ref{fig:TriangleZZcirc}.
There are nine distinct ways that the triangles can overlap, with two qubit gate depths given in the table on the right.
The depth 18 and 22 configurations can be avoided by ordering the up-pointing triangles (top: 0, bottom left: 1, bottom right: 2) and the down-pointing triangles (bottom: 2, top left: 0, top right: 1).
The worst case two qubit gate depth and two-qubit gate count per second-order Trotter step of $H_{\text{AFIM}}$ scales as,
\begin{equation}
    \text{\# 2q gates} \ = \ \frac{22}{3}N_S \quad , \quad \text{2q gate depth} \ = \ 17 \ . 
\end{equation}
We have used that there twice as many bonds as lattice sites on the kagome lattice, and are omitting boundary effects that only decrease the resource requirements.
This gate scaling should be compared to all-to-all connectivity where,
\begin{equation}
    \text{\# 2q gates} \ = \ 2N_S \quad , \quad \text{2q gate depth} \ = \ 4 \ . 
\end{equation}
The quantum simulations in Section~\ref{sec:qsims} employed three second order Trotter steps with an associated two-qubit depth (including evolution under $H_{SE}$) of $55$.

For completeness, we also provide quantum circuits that implement a second-order Trotter step of the AFHM Hamiltonian on the kagome lattice constrained to heavy-hex connectivity.
This requires evolution under $(X_iX_j+Y_iY_j+Z_iZ_j)$ for all pairs of nearest-neighbor sites.
Unlike in the AFIM, these terms do not commute on overlapping links, making an efficient second-order Trotterization more challenging.
Our strategy begins by partitioning the kagome lattice into up- and down-pointing triangles.
Evolution under the Heisenberg Hamiltonian is then performed for the three links on each triangle.
To obtain second-order Trotter accuracy, we use the symmetric triangle ordering (up-pointing($\delta_t/2$), down-pointing($\delta_t/2$), down-pointing($\delta_t/2$) reversed, up-pointing($\delta_t/2$) reversed).
As in the AFIM, the first step to implementing the evolution within a triangle is to act with a swap gate between a system qubit and the bath qubit, e.g. see Fig.~\ref{fig:TriangleZZcirc}a).
This reduces the problem to constructing a circuit that implements the Heisenberg evolution over three qubits with linear connectivity.

\begin{figure}[t]
\centering
\begin{minipage}{0.39\textwidth}
    \centering
    \includegraphics[width=\textwidth]{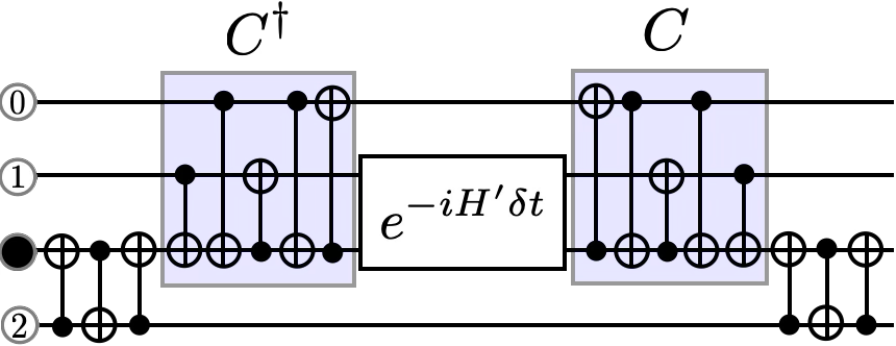}
\end{minipage}
\hfill
\begin{minipage}{0.6\textwidth}
    \centering
    \renewcommand{\arraystretch}{1.6}
    \begin{tabular}{|c||c|c|c|c|c|c|c|c|c|} \hline
    Shared triangle vertex&0 0&0 1&0 2&1 0&1 1&1 2&2 0&2 1&2 2 \\ \hline
    2q gate depth&26&27&31&27&28&32&31&32&36 \\ \hline
    \end{tabular}
\end{minipage}
\caption{Left: The decomposition of AFHM evolution within a triangle using the native gate set and connectivity of IBM's {\tt heron} quantum computers.
The evolution of $e^{-i H' \delta_t}$ can be synthesized with two two-qubit gates.
The vertices are labeled $0,1,2$ to track the orientation of each triangle (see Fig.~\ref{fig:TriangleZZcirc}).
Right: The upper row gives every combination of overlapping vertices between two triangles. 
The lower row gives the corresponding two-qubit gate depth.}
\label{fig:TriangleAFHMcirc}
\end{figure}

The authors of Ref.~\cite{Negishi:2026ujz} showed how to construct a circuit that performs this evolution exactly (without Trotter error).
Their construction applies a Clifford circuit $C$ that implements the following transformation,
\begin{align}
C^{\dagger}H C \ = \ H'
\end{align}
where,
\begin{align}
    H \ &= \ \left (X_0 X_1+Y_0 Y_1+Z_0 Z_1 \right )+\left (X_1 X_2+Y_1 Y_2+Z_1 Z_2 \right ) +\left (X_2 X_0+Y_2 Y_0+Z_2 Z_0 \right ) 
\end{align}
is the Heisenberg Hamiltonian over the three qubits on a triangle and 
\begin{align}
    H' \ &= \ \left (X_2+X_1+Z_1+Z_2-Z_1X_2-X_1Z_2+X_1X_2+Z_1Z_2+Y_1Y_2 \right ) \ .
\end{align}
The change of basis gives
\begin{align}
    e^{-i H \delta_t} \ = \ Ce^{-i H' \delta_t}C^{\dagger} \ ,
\end{align}
and the right-hand side can be synthesized exactly since it only involves Clifford gates and the two-qubit unitary $e^{-iH' \delta_t}$.
The latter can be exactly implemented with two two-qubit gates.
The full circuit is shown in Fig.~\ref{fig:TriangleAFHMcirc} and has a circuit depth of 18.

Since the evolution is exact, the symmetric ordering of triangles needed for a second-order Trotterization simplifies to (up-pointing($\delta_t/2$), down-pointing($\delta_t$), up-pointing($\delta_t/2$)).
The circuit depth can be reduced by optimally ordering the orientation of each triangle.
The depth for each configuration of overlapping triangle vertices is given in the right of Fig.~\ref{fig:TriangleAFHMcirc}.
The depth 32 and 36 configurations can be avoided by orienting the up-triangle (top: 0, bottom-left: 1, bottom-right: 2), down-triangle (bottom: 2, top-left: 0, top-right: 1) and up-triangle (top: 0, bottom-left: 1, bottom-right: 2).
This ordering gives,
\begin{equation}
    \text{\# 2q gates} \ = \ 18N_S \quad , \quad \text{2q gate depth} \ = \ 44  
\end{equation}
per second-order Trotter step of the AFHM evolution.
We emphasize that the exact synthesis of each triangle substantially reduces both Trotter error and circuit depth compared to a decomposition based on individual Heisenberg links.
Compared to the AFIM, Trotterized evolution in the AFHM requires approximately $2.5\times$ more two-qubit gates and depth.

\section{The steady state in the presence of depolarizing noise}
\label{app:noiseEffect}
\noindent
The quantum simulations in Section~\ref{sec:qsims} show that a steady state is reached in the presence of device noise.
Here we analyze how the steady state changes under two competing quantum channels (thermalizing and noisy) with different fixed points.
The scaling of the fixed point error is shown to agree with numerical simulations of noisy dissipative preparation of the ground state ($\beta=\infty$), as well as  with the results from {\tt ibm\_boston}.
We start by considering the simplest case of weak global depolarizing noise that occurs after each application of the dissipative quantum channel.
The scaling of the steady state error agrees with predictions from a more realistic noise model that is discussed below.

A global depolarizing channel $\Phi_d$ with strength $p$ acts on any density matrix as,
\begin{align}
\Phi_d(\rho) \ = \ (1-p)\rho \ + \ p \frac{\mathds{1}}{2^{N_S}} \ . 
\end{align}
The steady state $\rho_{\text{s.s.}}$ of the combined channels is defined as,
\begin{align}
\rho_{\text{s.s.}} \ &= \ (\Phi_d\circ \Phi)(\rho_{\text{s.s.}}) \nonumber \\
&= \ (1-p)\Phi(\rho_{\text{s.s.}}) \ + \ p \frac{\mathds{1}}{2^{N_S}} \ . 
\end{align}
Expanding $\rho_{\text{s.s.}} = \tilde{\rho}_S(\beta)+\delta\rho$ gives,
\begin{align}
\delta\rho \ &= \ (1-p)\Phi(\tilde{\rho}_S(\beta)+\delta \rho)\ + \ p \frac{\mathds{1}}{2^{N_S}} -\tilde{\rho}_S(\beta)\nonumber \\
&=\ (1-p)\left (\tilde{\rho}_S(\beta)+\Phi(\delta \rho)\right )\ + \ p \frac{\mathds{1}}{2^{N_S}} -\tilde{\rho}_S(\beta) \nonumber \\
&= \ p\left ( \frac{\mathds{1}}{2^{N_S}} -\tilde{\rho}_S(\beta) \right ) \ + \ \Phi(\delta \rho) \ + \ {\mathcal O}(p^2) \ . 
\label{eq:drho}
\end{align}
The second line has used the linearity of $\Phi$ as well as its fixed point, and the third line drops the higher order $\mathcal{O}(p\, \delta \rho)$ terms.

Next, define the right-eigenvectors of $\Phi$ as,
\begin{align}
    \Phi(v_k) \ = \ \lambda_k v_k \ .
\end{align}
The eigenvalues are real if $\Phi$ satisfies quantum detailed balance and have $|\lambda_k|\leq1$ with the steady state $v_0=\tilde{\rho}_S(\beta)$ and $\lambda_0=1$.
The left eigenvectors are defined as $w_k$ and the orthonormality condition is
\begin{align}
\text{Tr}\left (w_j^{\dagger } v_k\right ) \ = \ \delta_{jk} \ . 
\end{align}
Probability conservation gives $\text{Tr}(v_0) = \text{Tr}(\tilde{\rho}_S(\beta)) =1$ which implies $w_0 = \mathds{1}$ and $\text{Tr}(v_k)=0$ for $k> 0$.
Expressing,
\begin{align}
\delta \rho=\sum_k c_k v_k \quad , \quad \left ( \frac{\mathds{1}}{2^{N_S}} -\tilde{\rho}_S(\beta) \right )=\sum_k a_k v_k
\end{align}
gives,
\begin{align}
\delta \rho - \Phi(\delta\rho) \ &= \ p\left ( \frac{\mathds{1}}{2^{N_S}} -\tilde{\rho}_S(\beta) \right ) \nonumber \\
\sum_{k>0} c_k (1-\lambda_k)v_k \ &= \ p \sum_{k>0} a_k v_k \ ,
\end{align}
where the $k=0$ terms do not contribute because both sides are traceless.
Equating the terms gives $c_k = p a_k/(1-\lambda_k)$ and
\begin{align}
\delta \rho \ = \ p\sum_{k>0}  \frac{a_k}{1-\lambda_k} v_k \quad , \quad a_k \ = \ \text{Tr}\left [w_k^{\dagger} \left ( \frac{\mathds{1}}{2^{N_S}} -\tilde{\rho}_S(\beta) \right ) \right ] \ .
\end{align}
The size of $\delta \rho$ can be bounded by the spectral gap using $1-\lambda_k \geq (\lambda_0 - \text{max}_{k>0}|\lambda_k|) \equiv \Delta$,
\begin{align}
||\delta \rho||_1 \ \leq \ \frac{p}{\Delta} \bigg |\bigg|\left ( \frac{\mathds{1}}{2^{N_S}} -\tilde{\rho}_S(\beta) \right )\bigg |\bigg|_1  \ .
\label{eq:DepolarizingError}
\end{align}
Parametrically, the mixing time scales as $\tau_{\text{mix}} \sim 1/\Delta$ so $||\delta \rho||_1$ scales linearly with the mixing time,
\begin{align}
||\rho_{\text{s.s.}}-\tilde{\rho}_{S}(\beta)||_1 
\lesssim p \, \tau_{\text{mix}} \, \bigg |\bigg|\left ( \frac{\mathds{1}}{2^{N_S}} -\tilde{\rho}_S(\beta) \right )\bigg |\bigg|_1 \ .
\label{eq:DepolarizingError2}
\end{align}

Next, consider a more realistic scenario in which local depolarizing noise acts continuously during the system–environment Hamiltonian evolution. Specifically, consider the quantum channel
\begin{equation}
\label{Eq:de_noise}
    \Phi(\rho_S) =\Tr_E\left[{{\cal T}}\exp\left ({\int_0^T\mathcal{L}_N(t)dt}\right )(\rho_S \otimes \rho_E)\right] \ .
\end{equation}
where $\mathcal{L}_N$ is the system-bath Hamiltonian evolution together with the local depolarizing noise $\mathcal{L}_d(\rho)=\sum_i\left(\left({X_i \rho X_i+Y_i \rho Y_i+Z_i \rho Z_i}\right)/{3}-\rho\right)$ with strength $p_1$:
\[
\frac{\mathrm{d}\rho}{\mathrm{d}t}=\mathcal{L}_N(t){[\rho]}=-i[H(t),\rho]+p_1\mathcal{L}_d(\rho)
\]
A straightforward eigenvalue perturbation analysis would suggest that the steady-state error scales extensively with the system size, i.e., as $(N_S \,p_1)$. 
However, rapidly mixing Lindbladians have been shown to be stable against noise. 
Below, we summarize the main result of \cite{cubitt2015stability}:

\begin{theorem}[Stability of rapidly mixing local Lindbladians; Informal \cite{cubitt2015stability}]
Let $\mathcal{L} = \sum_X \mathcal{L}_X$ be a local Lindbladian acting on a quantum lattice system, where each term $\mathcal{L}_X$ has finite support and bounded strength. Assume that:
\begin{enumerate}
    \item $\mathcal{L}$ has a unique steady state $\rho_\infty$,
    \item the dynamics generated by $\mathcal{L}$ is rapidly mixing, i.e., for any initial state $\rho$,
    \begin{align}
        \| e^{t\mathcal{L}}(\rho) - \rho_\infty \|_1 \leq C \, e^{-\gamma t} \, \mathrm{poly}(N_S),
    \end{align}
    with constants $C,\gamma > 0$ independent of system size $N_S$.
\end{enumerate}

Let $\tilde{\mathcal{L}} = \mathcal{L} + \mathcal{E}$ be a perturbation of $\mathcal{L}$ by local terms $\mathcal{E}=\sum_x \mathcal{E}_x$ where each local perturbation has strength $\left\|\mathcal{E}_x\right\| \leq p_1$. and let $\tilde{\rho}_\infty$ denote the steady state of $\tilde{\mathcal{L}}$.

Then for any local observable $O_A$ supported on a fixed region $A$, the difference between the ideal and perturbed evolutions satisfies
$$
\left|\operatorname{Tr}\left[O_A e^{t \mathcal{L}}(\rho)\right]-\operatorname{Tr}\left[O_A e^{t \widetilde{\mathcal{L}}}(\rho)\right]\right| \leq \operatorname{poly}(|A|)\left\|O_A\right\| \frac{p_1}{\gamma},
$$
uniformly for all times $t$.
Equivalently, for the steady states,
$$
\left|\operatorname{Tr}\left[O_A \rho_{\infty}\right]-\operatorname{Tr}\left[O_A \tilde{\rho}_{\infty}\right]\right| \leq \operatorname{poly}(|A|)\left\|O_A\right\| \frac{p_1}{\gamma}
$$
\end{theorem}

Although this stability result has only been established for Lindbladian dynamics, we conjecture that a similar statement also holds for quantum channels. 
Assume the quantum channel has a unique steady state, and is rapidly mixing, i.e.  
\[
\|\Phi^{N_{\text{resets}}}(\rho)-\rho_\infty\|_1\leq C e^{-\gamma N_{\text{resets}}} \mathrm{poly}(N_S)
\]
where $\rho$ is arbitrary and $\gamma \propto 1/\tau_{\text{mix}}$ is the convergence rate.\footnote{
This definition is equivalent to constant global fidelity requiring $N_{\text{resets}}={\cal O}(\log(N_S))$ or constant local fidelity requiring $N_{\text{resets}} = {\cal O}(1)$.}
Then, under local depolarizing noise of strength $p_1$, the steady-state error of local observables also obeys
\[
\left| \operatorname{Tr}\!\left[ O_A (\tilde{\rho}_\infty - \rho_\infty) \right] \right|
    =\mathcal{O}\left({p_1}\tau_{\text{mix}}\right) \ .
\]
This gives the same error scaling obtained in the simplified noise model above, Eq.~\eqref{eq:DepolarizingError2}, but now applied to expectation values of local observables.

\begin{figure}[htbp]
    \centering
    \subfloat{\includegraphics[width=0.495\textwidth]{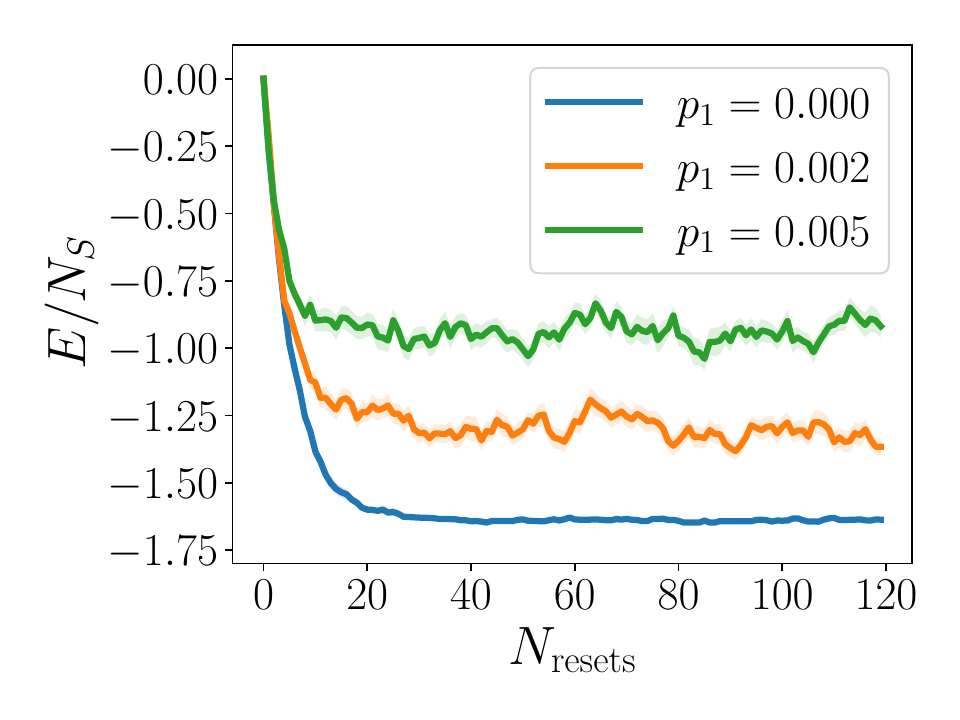}}
    \hfill
    \subfloat{\includegraphics[width=0.495\textwidth]{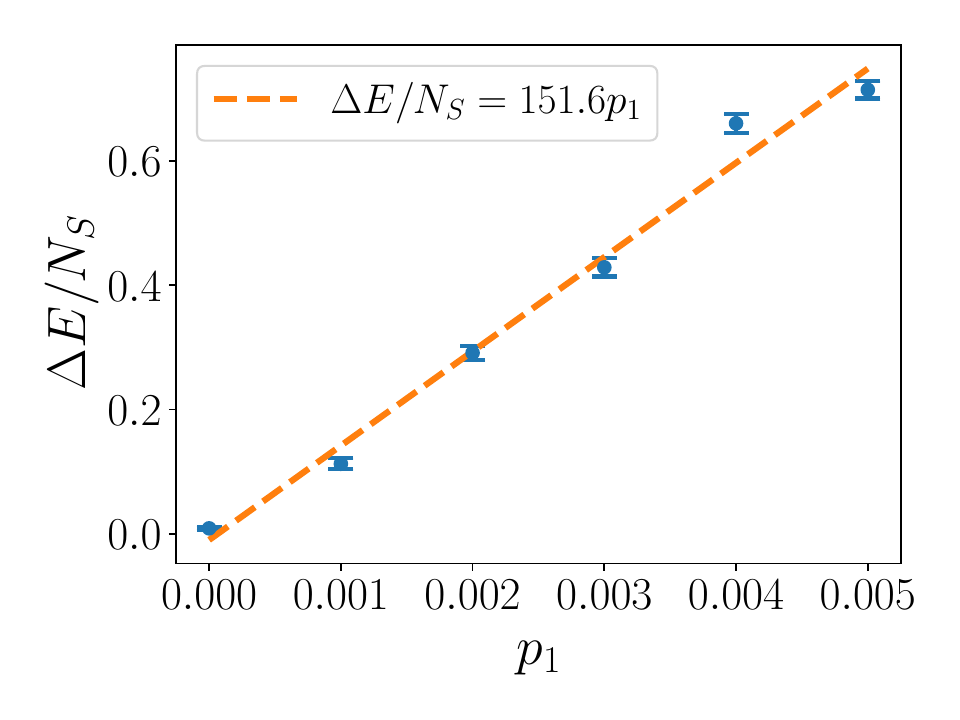}}
    \caption{Numerical simulation of dissipative ground-state preparation for the one-dimensional transverse-field Ising model $H=-J\sum_i Z_i Z_{i+1}+g\sum_i X_i$ with $N_S=8, N_E=4$ and $J=1, g=1.5$. Left: Energy density during the dissipative dynamics for different depolarizing noise strengths $p_1$. Stronger depolarizing noise leads to larger steady-state energy errors. Right: The energy error $\Delta E$ scales approximately linearly with the depolarizing noise strength.}
    \label{fig:noise_level_scaling}
\end{figure}

As shown numerically in Appendix~\ref{app:nQ_convergence}, the mixing time $\tau_{\text{mix}}$ scales approximately as $N_S/N_E$. Consequently
\begin{align}
\left| \operatorname{Tr}\!\left[ O_A (\tilde{\rho}_\infty - \rho_\infty) \right] \right|
    \ = \ \mathcal{O}\left(  \frac{p_1 N_S}{N_E}\right) \ ,
\label{eq:ErrorO}
\end{align}
To verify this scaling, we numerically investigate how the steady state energy density shifts when varying the depolarizing noise strength and the number of environment qubits.
The energy density is a local observable, so the deviation is expected to follow Eq.~\eqref{eq:ErrorO}.
First, we simulate dissipative preparation of the ground-state ($\beta=\infty$) for the one-dimensional transverse-field Ising model with $N_S=8$ and $N_E=4$ under different depolarizing noise strengths. 
The noisy Hamiltonian evolution is simulated using Trotterization of Eq.~\eqref{Eq:de_noise}.
The results, shown in Fig.~\ref{fig:noise_level_scaling}, demonstrate that stronger noise leads to larger energy errors. 
The deviation from the noiseless steady state energy $\Delta E$  scales approximately linearly with the depolarizing noise strength, consistent with Eq.~\eqref{eq:ErrorO}.

\begin{figure}[htbp]
    \centering
    \subfloat{\includegraphics[width=0.495\textwidth]{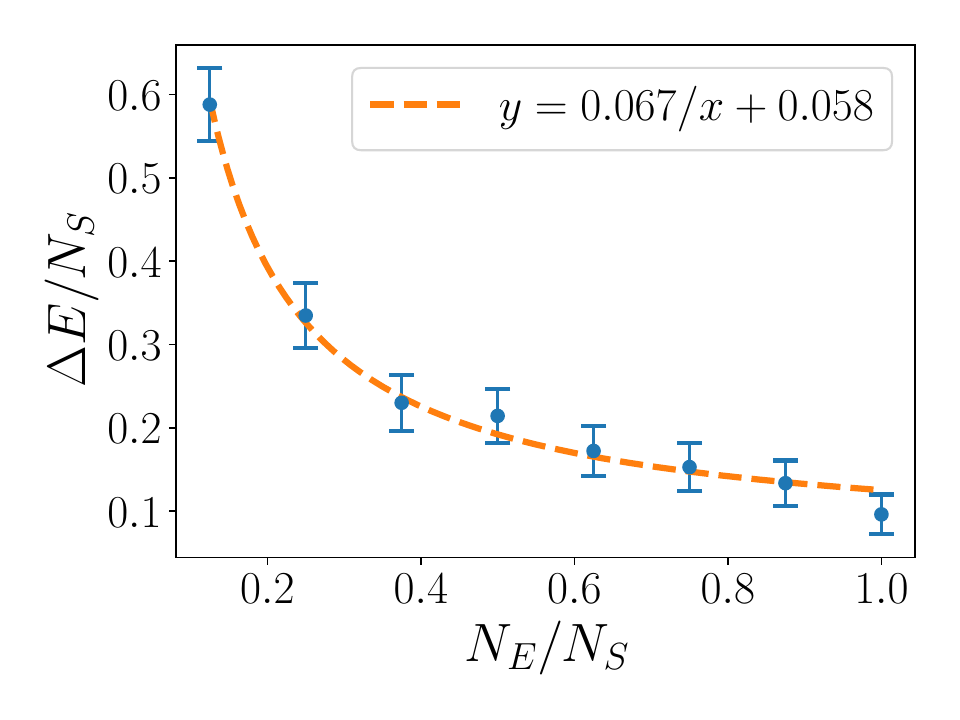}}
    \hfill
    \subfloat{\includegraphics[width=0.495\textwidth]{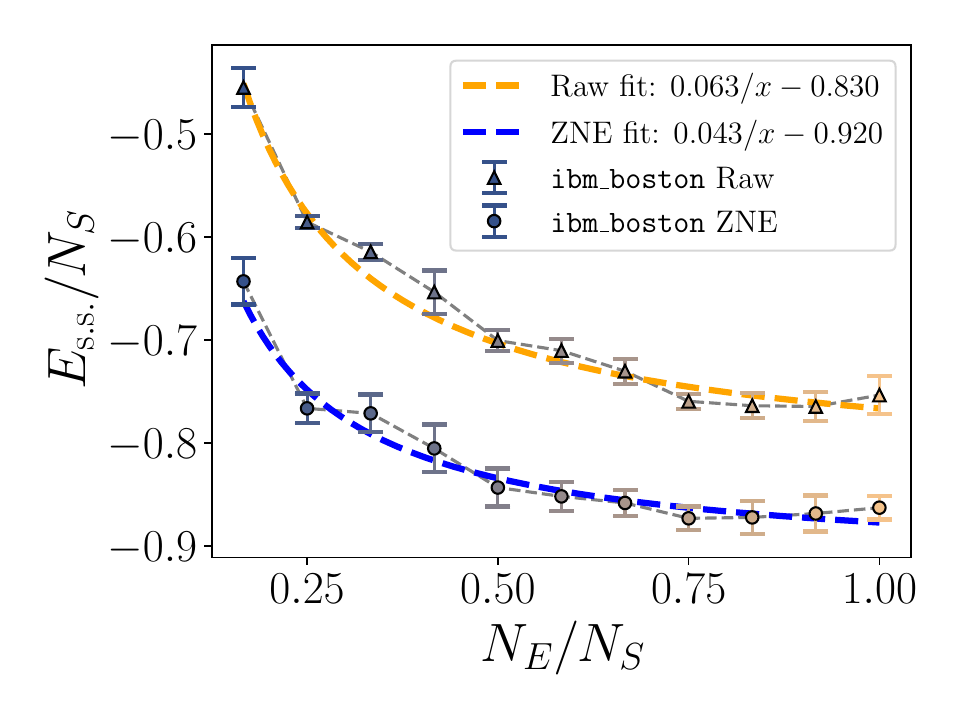}}
    \caption{Scaling of the steady-state energy error with the number of environment qubits $N_E$. 
    Left: Numerical simulation of dissipative ground-state preparation for the one-dimensional transverse-field Ising model with $N_S=8$ and depolarizing noise strength $p_1=0.002$. Right: Scaling of the $\beta=\infty$ steady state energy $E_{\text{s.s.}}$ obtained for $N_S=12$ in the AFIM on the kagome lattice using {\tt ibm\_boston}.
    Both the raw results and those obtained after ZNE are shown.}
    \label{fig:ancilla_qubits}
\end{figure}

Next, we simulate dissipative preparation of the ground-state ($\beta=\infty$) for the one-dimensional transverse-field Ising model varying $N_E$ with the depolarizing noise strength held fixed to $p_1= 2\times10^{-3}$. 
The numerical results, shown in the left panel of Fig.~\ref{fig:ancilla_qubits}, indicate that the deviation in the steady state energy decreases approximately inversely with $N_E$, in agreement with Eq.~\eqref{eq:ErrorO}. 
The right panel of Fig.~\ref{fig:ancilla_qubits} presents experimental results obtained from {\tt ibm\_boston} with $N_S=12$ and varying number of $N_E$.\footnote{The steady state energy reported here is larger than that presented in the main text.
This is because the energies in Fig.~\ref{fig:ancilla_qubits} do not employ contextual gate calibration, leakage post selection or idle qubit twirling, and also use the noisier {\tt reset\_2}. } 
The steady state energy $E_{\text{s.s.}}$ is obtained by averaging the energy over three values of $N_{\text{resets}}$ where the energy density does not vary by more than $0.05$.
The number of resets required to reach the steady state increases for fewer $N_E$; in the noiseless case $\tau_{\text{mix}}\propto N_S/N_E$.
To account for this, the number of resets for $N_E=\{2,3,4,5,6,7,8,9,10,11,12\}$ is taken to be
\[
\{(20+n),(14+n),(10+n),(9+n),(7+n),(7+n),(6+n),(5+n),(5+n),(4+n),(5+n)\} \quad \text{for }n=0,1,2 \ .
\]
The observed trend in the steady state energy is consistent with $E_{\text{s.s.}}\propto c_0N_S/N_E + c_1$ as predicted by Eq.~\eqref{eq:ErrorO}.
Also shown are the energies obtained after zero noise extrapolation (ZNE). 
We observe that ZNE follows a similar trend to the raw energies and is not able to effectively mitigate the increased noise sensitivity coming from decreasing $N_E$.
The fit coefficient for the raw data is $c_0=0.063(3)$ and for ZNE is $c_0=0.043(3)$.
Assuming this scaling is independent of $N_S$, then the difference between the ZNE $E_{\text{s.s.}}$ for $N_S = 12$ and $N_S=79$ would be predicted to be $\Delta E = 0.043 \times(79/60) = 0.06$.
This agrees reasonably well with the observed energy difference reported in Table~\ref{tab:energies} of $0.94(1)-0.86(1) = 0.08(1)$.
We remark that the $N_E=2$ simulations utilized up to $N_{\text{resets}}=22$ with raw (ZNE) two-qubit gate depths of $1063$ (1910).
This highlights the inherent noise robustness of dissipative thermal state preparation.
%

\section{Layout of the environment qubits on {\tt ibm\_boston}}
\label{app:bigKagome}
\begin{figure}
    \centering
    \includegraphics[width=\linewidth]{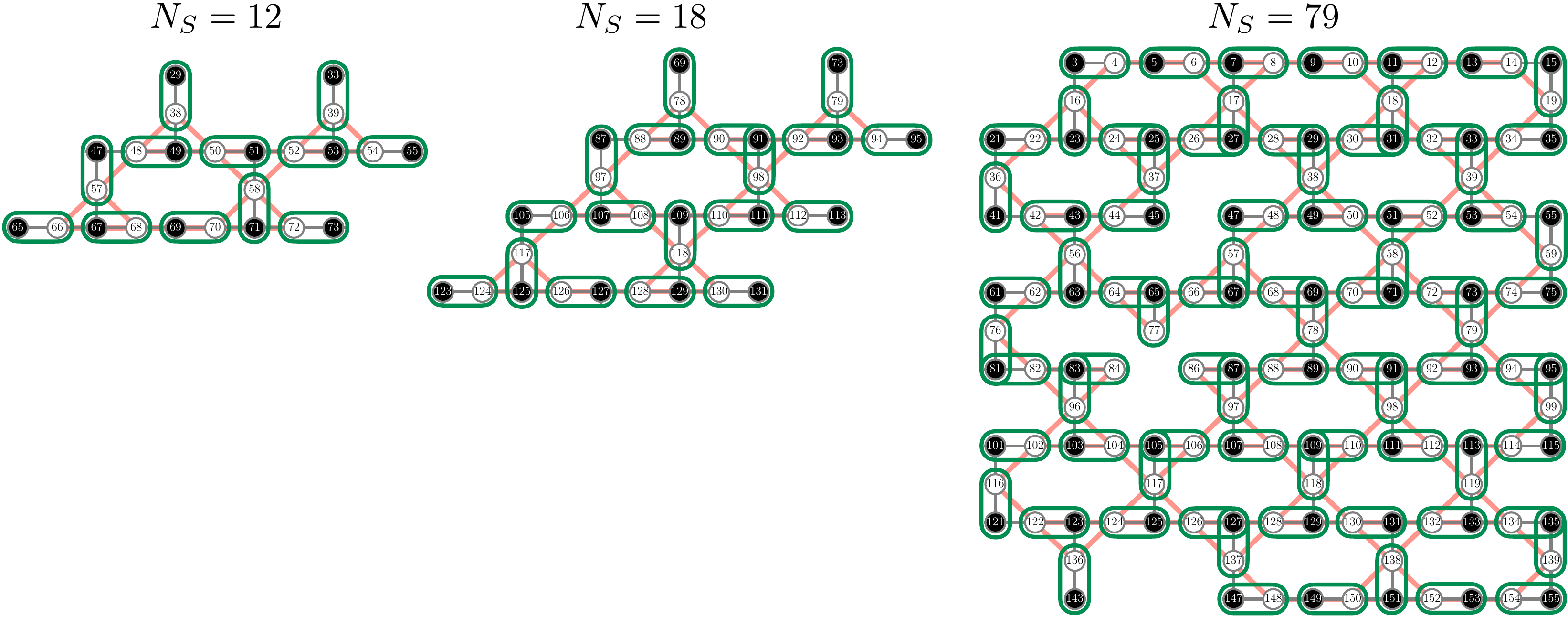}
    \caption{The pairing of system (white) and environment (black) qubits on the $N_S=12,18,79$ lattices that were used in the quantum simulations performed on {\tt ibm\_boston}. 
    The qubits numbering matches that on IBM's quantum cloud.
    The green chain-links represent the coupling between system and environment qubits.}
    \label{fig:IBM_lattice_SE}
\end{figure}

\noindent
The pairing of system and environment qubits that is used in the quantum simulations performed on {\tt ibm\_boston} in Section~\ref{sec:qsims} are shown in Fig.~\ref{fig:IBM_lattice_SE}. 
The green links show the system and environment qubits that are paired together. 
These layouts have $(N_S,N_E)=(12,12),(18,17),(79,60)$.
For environment qubits that can be paired with two system qubits, the system qubit is chosen randomly with uniform probability at each reset cycle.
The density of environment qubits is lower in the bulk of the $N_S=79$ lattice which locally decreases the strength of the dissipative channel.
In the absence of noise, this likely would not effect the steady state that is reached.
However, as shown in Appendix~\ref{app:noiseEffect}, the steady state is more impacted by noise when the density of environment qubits is lower.
This causes the bulk of the $N_S=79$ lattice to have higher energy density than on the boundary, i.e. effectively be at a higher temperature.
The gradient in energy density helps explain Fig.~\ref{fig:IBM_results_Energy_nE}{\bf b}, where observables averaged over the bulk were compared to QMC calculations at a $\beta_{\text{eff}}$ tuned to reproduce the total energy measured on {\tt ibm\_boston}.
While the bulk magnetization, $M_Z^{(\text{bulk})}$ and $M_X^{(\text{bulk})}$, are consistent with QMC, the connected correlator $\triangle_{ZZ}^{(\text{bulk})}$  are smaller in magnitude.
The average energy density in the bulk gets contributions from all three of these observables and is therefore higher than the average, a consequence of the lower density of environment qubits.
%

\section{Error mitigation}
\label{app:errormitigation}
\noindent
The error mitigation strategy used to obtain the results from {\tt ibm\_boston} reported in Section~\ref{sec:qsims} has several stages.
First, the real-time calibration data available on IBM's cloud is used to identify two-qubit gates with anomalously larger error rates $\epsilon_{2q}>0.05$.
The lattice-to-qubit mapping is chosen to avoid these two qubit gates.
For $N_S=79$ this results in cutting out one system qubit and one environment qubit as shown in Fig.~\ref{fig:IBM_lattice_SE}.
Next, the quantum circuits are compiled using a $XX$ dynamical decoupling sequence to mitigate qubit idling errors~\cite{Viola:1998jx,Ezzell_2023}.
Additionally, all native two-qubit gates ($CZ$ and $R_{ZZ}(\theta)$) are Pauli twirled.
The $CZ$ gate is Clifford so it is twirled by all 16 combinations of Paulis appended to the input and outputs legs that leaves the gate invariant.
The $R_{ZZ}(\theta)$ gate is not Clifford and is only twirled by the 8 combinations of Paulis that leave the gate invariant~\cite{Farrell:2025nkx,Kim:2021gvc}.
The idle periods on the system qubits during {\tt reset} instructions on environment qubits (duration 2232~ns) are Pauli twirled as well.
This partial twirling does not guarantee a stochastic Pauli channel, but has empirically been found to reduce coherent errors.
Measurement errors are mitigated by enabling {\tt measurement\_twirling} in Qiskit's sampler primitive.
This randomly applies $X$ gates to the qubits before measurements to remove measurement bias.
These bit flips are removed in classical post-processing. 
Lastly, measurement bitstrings are postselected against the detection of leakage outside of the computational subspace~\cite{Kim2026InPrep}. 
Postselection survival fidelities were 89\% for the $N_S=12$, 85\% for $N_S=18$, and 22\% for $N_S=79$.

We initially tried using operator decoherence renormalization (ODR)~\cite{Farrell:2023fgd,Urbanek:2021oej,ARahman:2022tkr} to mitigate errors.
In ODR, additional reference circuits are run with a similar structure as the physics circuit but where the output expectation values are known.
The deviation of measured observables from their known values in the reference circuit is then used to rescale measured observables in the physics circuit.
For a reference circuits, we set the 3 Trotter steps in our circuit to have time steps $[\delta_t,0 ,-\delta_t]$.
In the absence of errors, the Trotter steps cancel and implement $\mathds{1}$.
A successful implementation of ODR relies on the assumption that noise affects the reference and physics circuits similarly. 
In dissipative dynamics, however, this assumption breaks down because the system and environment qubits are decoupled in the reference circuit but are coupled in the physics circuit. 
As a result, the system qubits in the reference circuit continuously decohere, whereas in the physics circuit they become entangled with the environment through the application of $U(T)$. 
Measurements of the environment qubits in the physics circuit therefore partially counteract decoherence on the system qubits.

\subsection{Reset-error adjusted \texorpdfstring{$\beta^*$}{}}
\noindent
One source of error comes from imperfect qubit reset. 
We will show that this changes the local temperature of each environment qubit resulting in a reset-error adjusted $\beta^*$ that is lower than the target $\beta$.

Define confusion matrices $\hat{X}_R$ and $\hat{X}_M$ that relate the input ${\vec I}$ and output ${\vec O}$ probability distributions after reset and measurement by
\begin{align}
\vec{O} \ = \ \hat{X}. \vec{I} \ .
\end{align}
The entries of $\hat{X}_M$ are obtained by assuming no state preparation error and recording the measurement probabilities of a qubit prepared in $|0\rangle$ or $|1\rangle$.
Preparing the qubit in $|0\rangle$ or $|1\rangle$, applying {\tt reset}, and then measuring gives the entries of $\hat{X}_M.\hat{X}_R$.
The entries of $\hat{X}_R$ are obtained by multiplying by $(\hat{X}_M)^{-1}$.
The reset confusion matrix has entries
\begin{align}
\hat{X}_R \ = \
\begin{pmatrix}
c_{00} & c_{01} \\
1-c_{00} & 1-c_{01}
\end{pmatrix}
\label{eq:XR}
\end{align}
where probability conservation requires the columns to sum to one. 
A perfect reset has $c_{00}=c_{01}=1$.

The input probabilities $\vec{I}$ before reset are needed to determine the state after reset.
In the steady state, the bath qubit probabilities do not change under the application of $\Phi$.
Therefore, the input probabilities can be determined by reading out the bath qubits in the final reset cycle,  and then applying $(\hat{X}_M)^{-1}$ to get $\vec{I}$.
The state after reset is ${\vec O} = (O_0,O_1)^T=\hat{X}_R. \vec{I}$.
To sample from the environment Gibbs state, each qubit is acted on by $X_{i_E}$ with probability given in Eq.~\eqref{eq:pBeta_E}.
The probability of the qubit being in $|1\rangle$ is then,
\begin{align}
\text{Pr}(|1\rangle) \ &= \ \frac{1}{\omega_{\text{max}}}\int_0^{\omega_{\text{max}}} d\omega \frac{O_0 e^{-\beta \omega/2} \ + \ O_1 e^{\beta \omega/2}}{2 \cosh(\beta \omega/2)} \nonumber \\
& = \frac{1}{\beta \omega_{\text{max}}}\left [O_0\beta \omega_{\text{max}} \ + \ (O_1 - O_0)(\log \left (1+e^{\beta \omega_{\text{max}}} \right ) - \log 2) \right ] 
\ ,
\label{eq:Pr1}
\end{align}
where we have averaged over the uniform random Bohr frequency $\omega$.
The probability of a qubit being $|1\rangle$ at an effective temperature $\beta^*$ is,
\begin{align}
\text{Pr}(|1\rangle) \ &= \  \frac{1}{\omega_{\text{max}}}\int_0^{\omega_{\text{max}}} d\omega\frac{ e^{-\beta^* \omega/2}}{2 \cosh(\beta^* \omega/2)} \nonumber \\
& = \ \frac{1}{\beta^* \omega_{\text{max}}}\left [\log 2 - \log \left (1+e^{-\beta^* \omega_{\text{max}}} \right ) \right ]  \ .
\label{eq:Pr1betas}
\end{align}
Equating Eq.~\eqref{eq:Pr1} and Eq.~\eqref{eq:Pr1betas} gives a transcendental equation for $\beta^*$ that can be solved numerically.
For $\beta=\infty$, Eq.~\eqref{eq:Pr1} simplifies and gives the maximum inverse-temperature $\beta^{\text{max}}$ that can be reached,
\begin{align}
O_1 \ = \ \frac{1}{\beta^{\text{max}} \omega_{\text{max}}}\left [\log 2 - \log \left (1+e^{-\beta^{\text{max}} \omega_{\text{max}}} \right ) \right ] \ .
\end{align}
For $\beta^{\text{max}} \omega_{\text{max}} \gg 1$ this has the approximate solution,
\begin{align}
\beta^{\text{max}} \ \approx \ \frac{\log2}{O_1 \omega_{\text{max}}}\left (1 \ + \ {\cal O}(e^{-\log 2/O_1}) \right ) \ .
\end{align}
\begin{figure}
    \centering
    \includegraphics[width=0.75\linewidth]{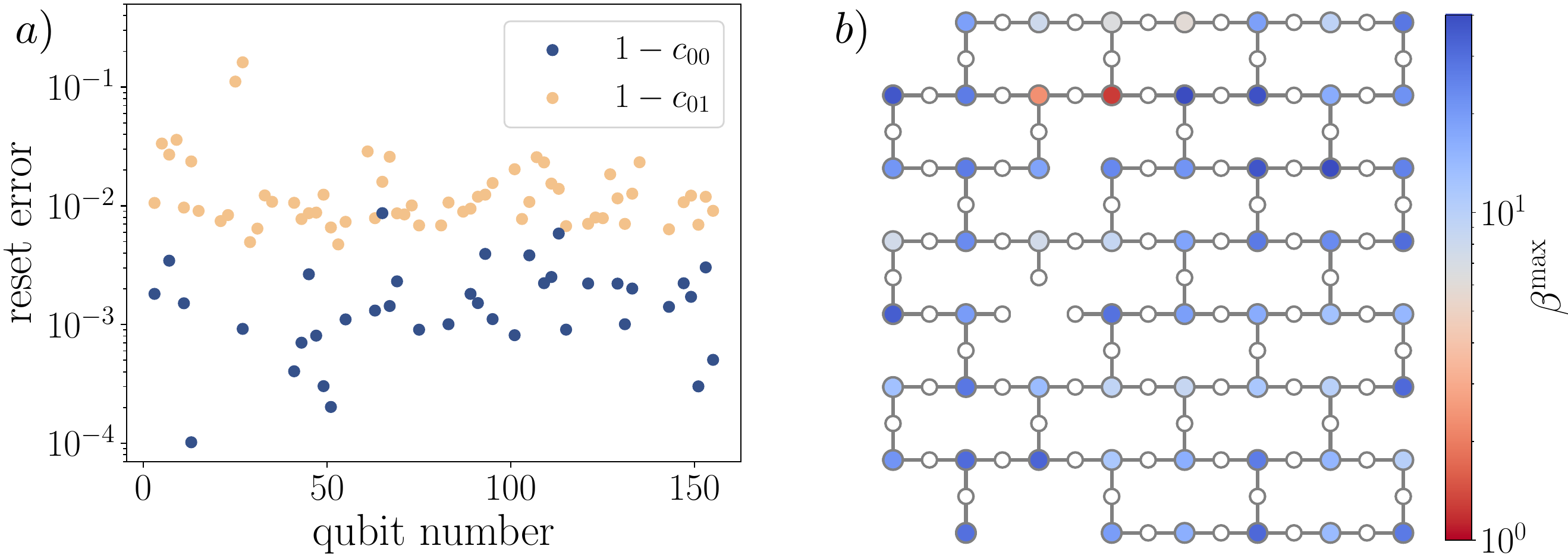}
    \caption{a) The entries in the reset confusion matrix $\hat{X}_R$, defined in Eq.~\eqref{eq:XR}, for the environment qubits used on the $N_S=79,N_E=60$ kagome lattice.
    The qubit labeling matches the IBM cloud, see Fig.~\ref{fig:IBM_lattice_SE}.
    b) The maximum inverse-temperature $\beta^{\text{max}}$ that can be reached in the presence of reset errors for the environment qubits used on the $N_S=79,N_E=60$ lattice. 
    The white colored qubits are the system qubits.}
    \label{fig:reset_error}
\end{figure}
The entries of the reset confusion matrix $\hat{X}_R$ for the environment qubits used on the $N_S=79$ lattice are shown in Fig.~\ref{fig:reset_error}a).
Most values are between $10^{-4}$ and $5\times10^{-2}$ with two outlier on qubits $q_{25}$ and $q_{27}$.
The corresponding $\beta^{\text{max}}$ is shown in Fig.~\ref{fig:reset_error}b) overlaid on the $N_S=79,N_E=60$ lattice.
Most qubits have $\beta^{\text{max}}\approx 19$,
with the two outliers having 
$\beta^{\text{max}} = 1.25$ and $\beta^{\text{max}} = 2.32$ for $q_{27}$ and $q_{25}$ respectively.
These hot bath qubits manifest in spatial variation of the observables measured on the system qubits.
The spatial median and minimum $\beta^*$ for all temperatures run on {\tt ibm\_boston} are given in Table~\ref{tab:betas}.
The hottest environment qubit is always $q_{27}$.
\begin{table}[t]
\centering
\renewcommand{\arraystretch}{1.4}
\begin{tabular}{|c||c|c|c|} 
\hline
$\beta$  & median($\beta^*$)  & min($\beta^*$) & $\beta_{\text{eff}}$   \\ 
\hline\hline
$\infty$ & 19.2 & 1.25 & 0.34(1)\\
\hline
0.5 & 0.49 & 0.36 & 0.21(1)\\
\hline
0.25 & 0.25 & 0.2 & 0.14(1) \\
\hline
\end{tabular}
\renewcommand{\arraystretch}{1.0}
\caption{The inverse temperature $\beta$ of the environment for the thermal states prepared on {\tt ibm\_boston} is given in column 1, the spatial median of $\beta^*$ in column 2, the smallest $\beta^*$ in column 3 and the effective temperature $\beta_{\text{eff}}$ in column 4.}
\label{tab:betas}
\end{table}
%

\subsection{Zero noise extrapolation}
\noindent
The final step of our error mitigation pipeline is Zero Noise Extrapolation (ZNE)~\cite{Li:2016vmf,Temme:2016vkz}, which was also used for circuits with mid-circuit measurements in Ref.~\cite{Shirgure:2026ywj}.
In our implementation, we apply a transformation pass that replaces $CZ\to CZ \cdot CZ \cdot CZ$ and $R_{ZZ}(\theta) \to R_{ZZ}(\theta)\cdot R_{ZZ}(-\theta)\cdot R_{ZZ}(\theta)$ with probability $p_{\text{ZNE}}$.
In addition, errors during system qubit idles while the environment qubits are reset are amplified by extending the idle duration.
In the absence of errors, these error amplification operations would do nothing.
Assuming the dominant source of errors are from two-qubit gates and qubit idling, this amplifies our error rate by an amount that is controlled by $p_{\text{ZNE}}$.
We run our circuits with $p_{\text{ZNE}}\in [0,0.1,0.2,0.35,0.5]$ and fit our measured observables to a single exponential,\footnote{For $N_{\text{resets}}<7$ and $\beta=1/2,1/4$ only three noise rates $p_{\text{ZNE}}\in [0,0.2,0.4]$ are used. For $N_{\text{resets}}=7$ we increased this to $p_{\text{ZNE}}\in [0,0.1,0.2,0.3,0.4]$ to decrease the error bars in the spatially resolved observables.}
\begin{align}
\langle O\rangle_{\text{meas}}(p_{\text{ZNE}}) \ = \ \langle O\rangle_{\text{ZNE}} \, e^{-\lambda \, (1+2p_{\text{ZNE}})}\ . 
\label{eq:OZNE}
\end{align}
Thus, our measurements give $5$ points to perform a two parameter fit of $\langle O\rangle_{\text{ZNE}}$ and $\lambda$.

\begin{figure}
    \centering
    \includegraphics[width=0.85\linewidth]{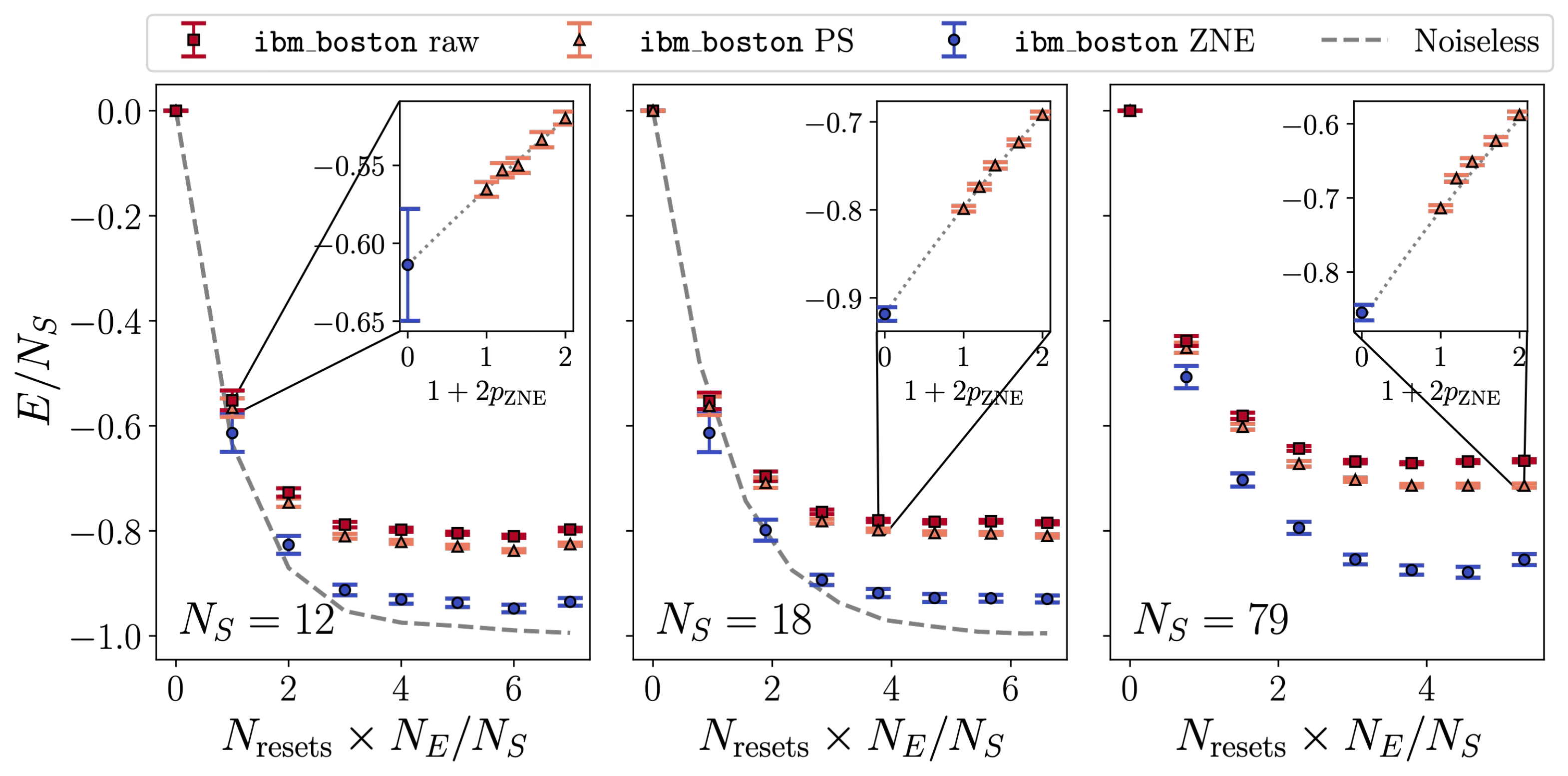}
    \caption{The energy densities measured on {\tt ibm\_boston} for $\beta=\infty$ and system sizes $N_S=12$ (left), $N_S=18$ (center) and $N_S=79$ (right).
    The red squares are the raw results, orange triangles are after leakage post-selection and the blue circles are the ZNE extrapolated value.
    The inset shows the energy densities obtained at different noise amplification rates. 
    The gray dotted line is the extrapolation to zero noise.
    The noiseless gray dashed curve is from statevector simulations.}
    \label{fig:ZNE_extrapolation}
\end{figure}
The raw results for the measured energy density at $\beta=\infty$ are compared to those after leakage post-selection (PS) and ZNE extrapolation in Fig.~\ref{fig:ZNE_extrapolation}.
The improvement between raw and PS is modest, and is most significant for $N_S=79$.
The PS results have systematically higher energies than the ZNE extrapolation due to all Pauli observables being biased to the pure-noise value of $E=0$.
The insets give the energy density evaluated at the different $p_{\text{ZNE}}$, as well as the extrapolated fit.
The fit value of $\lambda$ at $N_{\text{resets}}=4$ is larger than $N_{\text{resets}}=1$ indicating the effect of noise is more non-linear due to the circuit being deeper.
However, the curvature remains roughly the same between $N_{\text{resets}}=4$ and $N_{\text{resets}}=7$ because the system has reached a steady state.
In all cases, the observables $\langle O\rangle_{\text{meas}}(p_{\text{ZNE}})$ scales nearly linearly with $p_{\text{ZNE}}$ indicating that the extrapolation is not very sensitive to the single-exponential functional form in Eq.~\eqref{eq:OZNE}.
The ZNE extrapolation predicts larger energies than the noiseless results for $N_{\text{resets}}>2$.
This could be because the mid-circuit measurements contribute significantly to the noise, but are not being amplified.
We have found that choosing a qubit layout that avoids two-qubit gates with high error rates is essential for ZNE to perform well.
This is likely because randomly amplifying the noise of gates is more accurate when the gates have similar error rates.

\begin{figure}
    \centering
    \includegraphics[width=\linewidth]{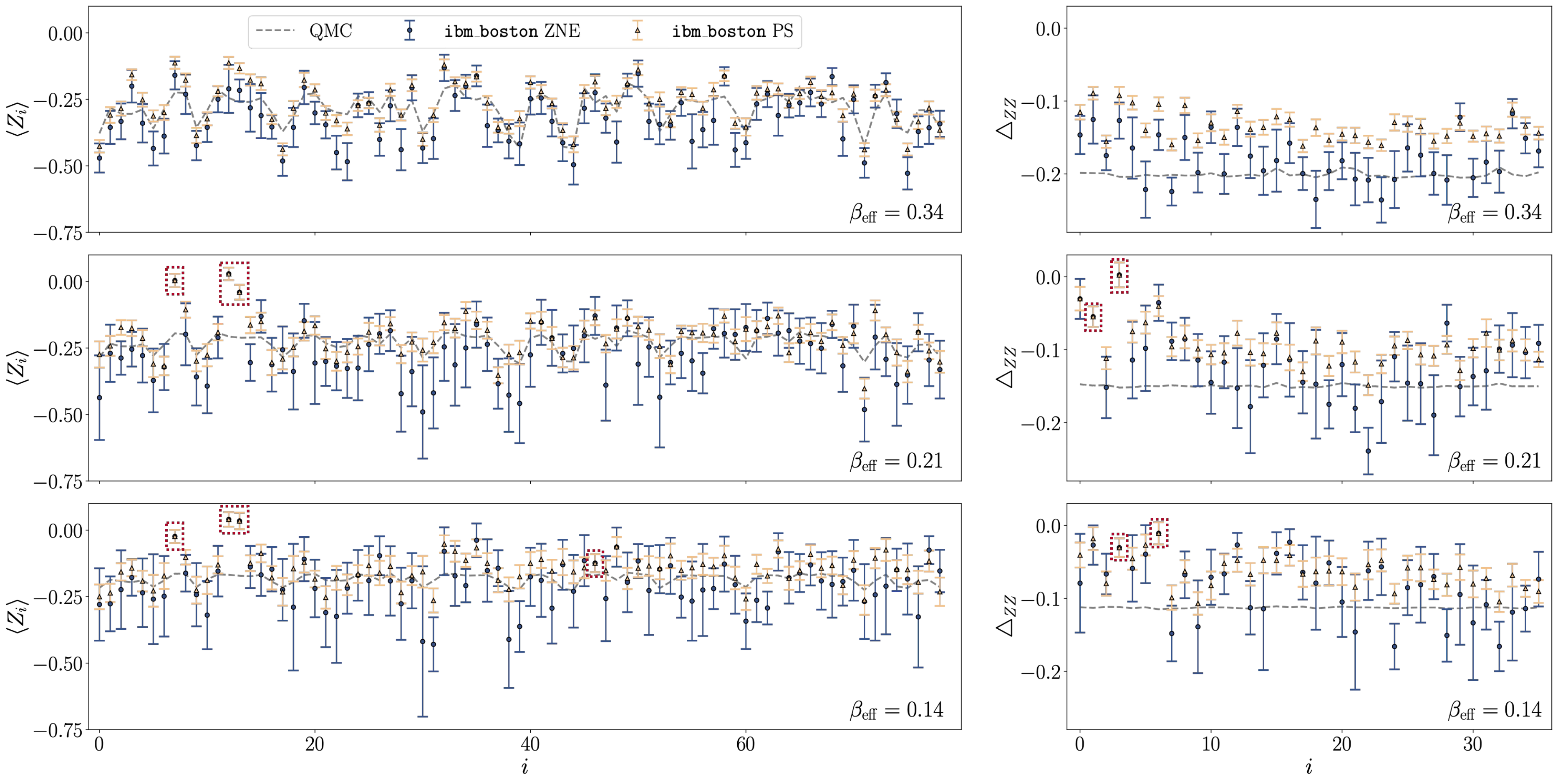}
    \caption{The spatially resolved magnetization (left) and triangle $ZZ$ correlator (right) measured on {\tt ibm\_boston} at $N_{\text{resets}}=7$ and compared to QMC evaluated at inverse-temperature $\beta_{\text{eff}}$.
    Lattice site and triangle labels are numbered left-to-right and top-to-bottom on the $N_S=79$ kagome lattice. 
    The quantities outlined in red do not have a ZNE extrapolation as they change sign under noise amplification.
    }
    \label{fig:ibm_QMC_compare}
\end{figure}
ZNE is also applied to the magnetization density $\langle Z_i\rangle$ and triangle correlator $\triangle_{ZZ}$ shown in Fig.~\ref{fig:IBM_results_Energy_nE}{\bf c} and {\bf d}.
The measured observables after leakage post-selection and ZNE are shown in Fig.~\ref{fig:ibm_QMC_compare} and compared to QMC evaluated at $\beta_{\text{eff}}$.
These observables have much larger error bars  than the energy since site-resolved observables have fewer statistics than spatially-averaged quantities.
The magnetization (left column) correctly resolves the spatial fluctuations at $\beta_{\text{eff}}=0.34$ that are due to the boundaries.
The error bars grow at higher temperature, but the magnetization is consistent with increasing spatial homogeneity that is seen in QMC.
The $\triangle_{ZZ}$ correlators (right column) measured on {\tt ibm\_boston} show significantly more spatial fluctuations than QMC for all temperatures.
One reason that the quantum data differs from QMC may be due to the density of bath qubits being larger on the boundaries of the lattice than in the bulk, see Fig.~\ref{fig:IBM_lattice_SE}.
This increases the strength of the thermalizing channel on the boundaries relative to the bulk.
A stronger dissipative channel is less sensitive to noise as discussed in Appendix~\ref{app:noiseEffect}, and reaches a steady state with a lower energy.
Therefore, the spins at the boundary are  locally at a lower temperature than the global average.
Another source of spatial variation comes from the environment qubits having different temperatures $\beta^*$ due to reset errors.
In particular, there are two hot environment qubits in the upper-middle region of the lattice as shown in Fig.~\ref{fig:reset_error}.
These spatial variations in the temperature affect $\triangle_{ZZ}$ more than $\langle Z_i\rangle$ as it is more sensitive to thermal fluctuations as shown in Fig.~\ref{fig:IBM_results_Energy_nE}{\bf b}.
Lastly, even in the absence of noise, there would still be deviations between the quantum simulation and QMC.
This is because the approximation of the dissipative channel $\Phi$ uses only three Trotter steps of time evolution, which gives rise to errors in the steady state that are especially significant when targeting lower temperatures.

The exponential ZNE fit in Eq.~\eqref{eq:OZNE} is sometimes unstable due to insufficient statistics in the site-resolved observables.
One source of instability occurs when the observable changes sign under the noise amplification.
This happens for the observables outlined with red boxes in Fig.~\ref{fig:ibm_QMC_compare} and is primarily localized around the hot environment qubits.
For these cases, we use the value of the observable after leakage post-selection and do not apply ZNE.
Additionally, there are some observables where the exponential ZNE fit has a large error compared to the standard deviation of the data after leakage post selection.
We use a cutoff of $\sigma_{\text{sd}}^{(\text{ZNE})} > 5\sigma_{\text{sd}}^{(\text{PS})}$, and in this case apply a linear extrapolation,
\begin{align}
\langle O\rangle_{\text{meas}}(p_{\text{ZNE}}) \ = \  \,  a(1+2p_{\text{ZNE}})+\langle O\rangle_{\text{ZNE}}\ . 
\label{eq:OZNE_linear}
\end{align}
Increasing the number of randomizations (currently 50) and shots-per-randomization (currently 200) would likely make this step unnecessary.

\section{Additional classical simulations}
\label{app:csim}
\begin{figure}
    \centering
    \includegraphics[width=\linewidth]{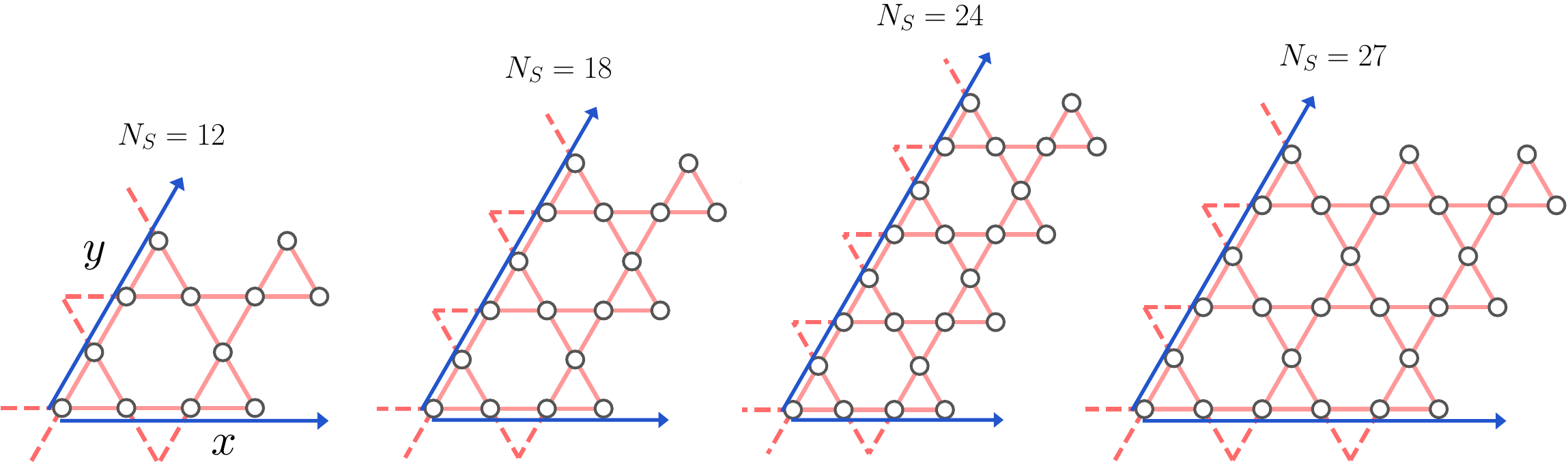}
    \caption{Kagome lattices with  $N_S=12,18,24,27$ sites corresponding to $(L_x,L_y)=(2,2),(2,3),(2,4),(3,3)$ unit cells.
    The dashed pink lines are additional connections present with PBCs.}
    \label{fig:kagomeLayouts}
\end{figure}

\noindent
In Section~\ref{sec:classicalsims}, simulations were performed on kagome lattice with $N_S=12,18,24,27$ sites 
corresponding to $(L_x,L_y)=(2,2),(2,3),(2,4),(3,3)$ unit cells.
The corresponding geometries are shown in Fig.~\ref{fig:kagomeLayouts}.
With PBCs, each site is connected to four other sites, with the connections that wrap around the lattice represented by the dashed pink lines.

\begin{figure}
    \centering
    \includegraphics[width=0.75\linewidth]{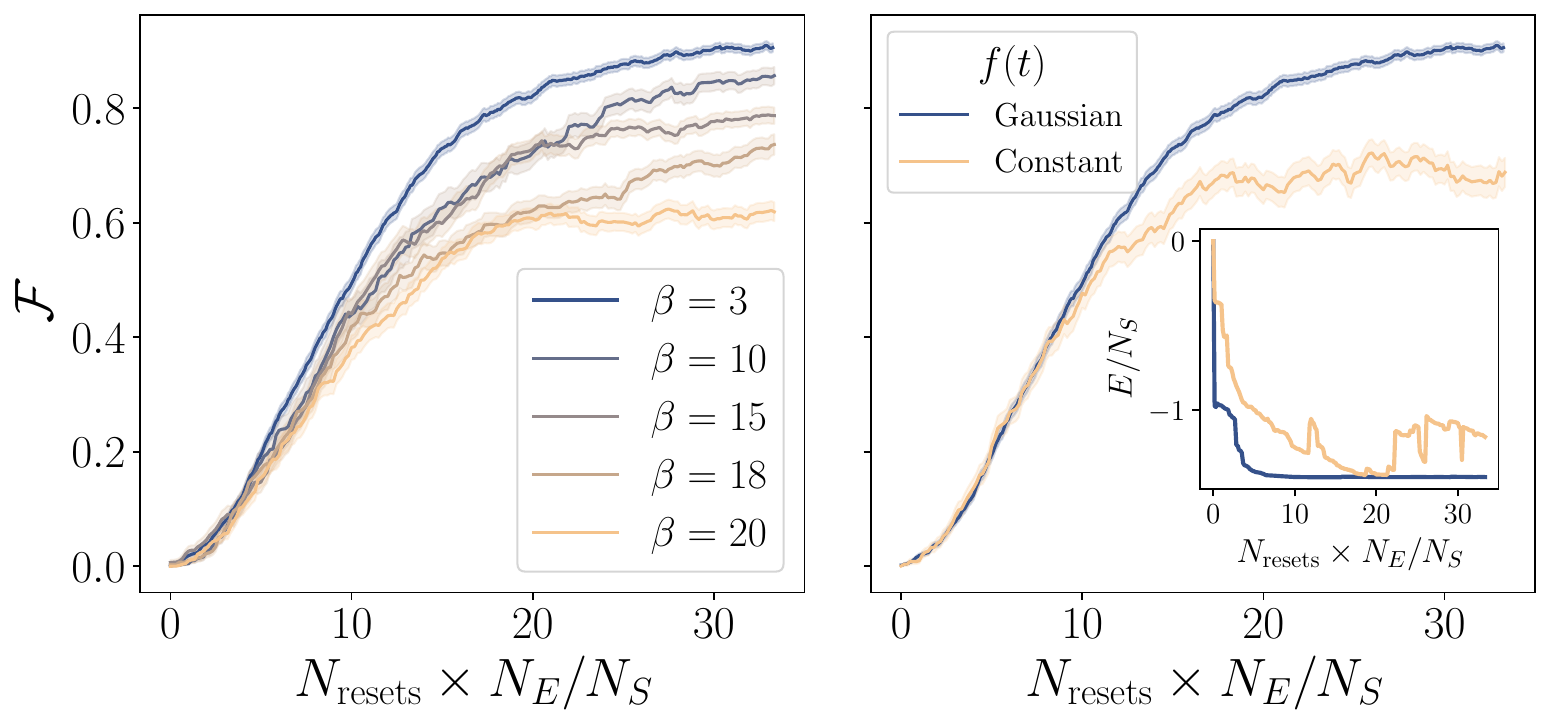}
    \caption{Left: Comparison of the rate of convergence to the exact thermal state with a range of inverse temperatures between $\beta=3$ and $\beta=20$.
    Right: Comparison of the rate of convergence to the thermal state using a Gaussian ($\sigma=1/4$) and constant ($\sigma=\infty$) filter function $f(t)$.
    The inset shows the energy over a single trajectory of the quantum channel.
    Both plots are on a $N_S=12$ kagome lattice with PBCs and have $T=8$. The remaining parameters are given in the caption of Fig.~\ref{fig:IsingHeisFid}.}
    \label{fig:LargeBeta_gaussian}
\end{figure}

In Fig.~\ref{fig:mix_scaling}{\bf b}, the mixing time in the AFIM is shown to have a local maximum at $\beta_c\approx1.4$, followed by a plateau for $2\leq\beta\leq3$.
The plateau is due to a hierarchy of energy scales in the spectrum, $\Delta E_{\text{non-ice}}/\Delta E_{\text{ice}}\gg 1$, as shown in Fig.~\ref{fig:mix_scaling}{\bf a}.
For,
\[
1/\Delta E_{\text{non-ice}}\ll\beta \ll1/\Delta E_{\text{ice}} \quad \longrightarrow \quad 0.6\ll\beta \ll10 \ ,
\] 
the Gibbs ensemble is close to a uniform distribution over the eigenstates in the ice manifold.
As a consequence, the Gibbs ensemble does not change for $2\leq \beta \leq 3$ and the mixing time remains constant.
The mixing time is expected to increase again for $\beta>10$ as the eigenstates in the ice manifold develop different Boltzmann weights.
We confirm this with additional classical simulations at $\beta=10,12,15,18,20$, with all other parameters held fixed.
The fidelity with the exact state is shown in the left plot of Fig.~\ref{fig:LargeBeta_gaussian} and compared to the convergence with $\beta=3$.
The convergence is slower for larger $\beta$ indicating a longer mixing time, as expected.
Notably, the fixed point error also increases for larger $\beta$, with a maximum fidelity of ${\cal F} \approx0.6$ for $\beta=20$.
This is because a constant fixed point error generally requires $T=\cal{O}(\beta)$ as discussed in Methods~\ref{methods:algo}.
Specifically, an evolution time $T\sim 1/\delta E$ is needed if the Boltzmann weights are non-negligible and vary in magnitude over an energy scale $\delta E$.
This occurs if there are energy eigenstates with energy $(E-E_0)\sim 1/\beta$.

One of the parameters that can be tuned in the dissipative thermal state preparation algorithm is the width $\sigma$ of the Gaussian filter function $f(t)$ defined in Eq.~\eqref{eq:f(t)}.
A larger $T/\sigma$ makes the time-dependent coupling and decoupling of the system to the environment smoother, decreasing the fixed point error.
The Gaussian width dependence is illustrated in the right plot of Fig.~\ref{fig:LargeBeta_gaussian} which compares the convergence of the fidelity for a Gaussian with $\sigma=1/4$ and a constant filter function ($\sigma=\infty$).
The inset shows the energy measured in a single trajectory of the quantum channel.
The constant filter function achieves a worse steady state fidelity, and the trajectory in energy show that this is due to large jumps in energy that are suppressed when using a Gaussian. 
One way to understand this is to consider acting the dissipative quantum channel with $\beta=\infty$ on the system initialized to the ground state.
With a Gaussian filter function, the interaction is slowly turned on and off.
Assuming no gap closing, this is adiabatic and preserves the ground state of the system.
By contrast, a constant filter function is not adiabatic, and the coupling between system and environment can transition the system to an excited state.
Similar intuition can likely be applied to finite $\beta$ as well.

\begin{figure}
    \centering
    \includegraphics[width=0.75\linewidth]{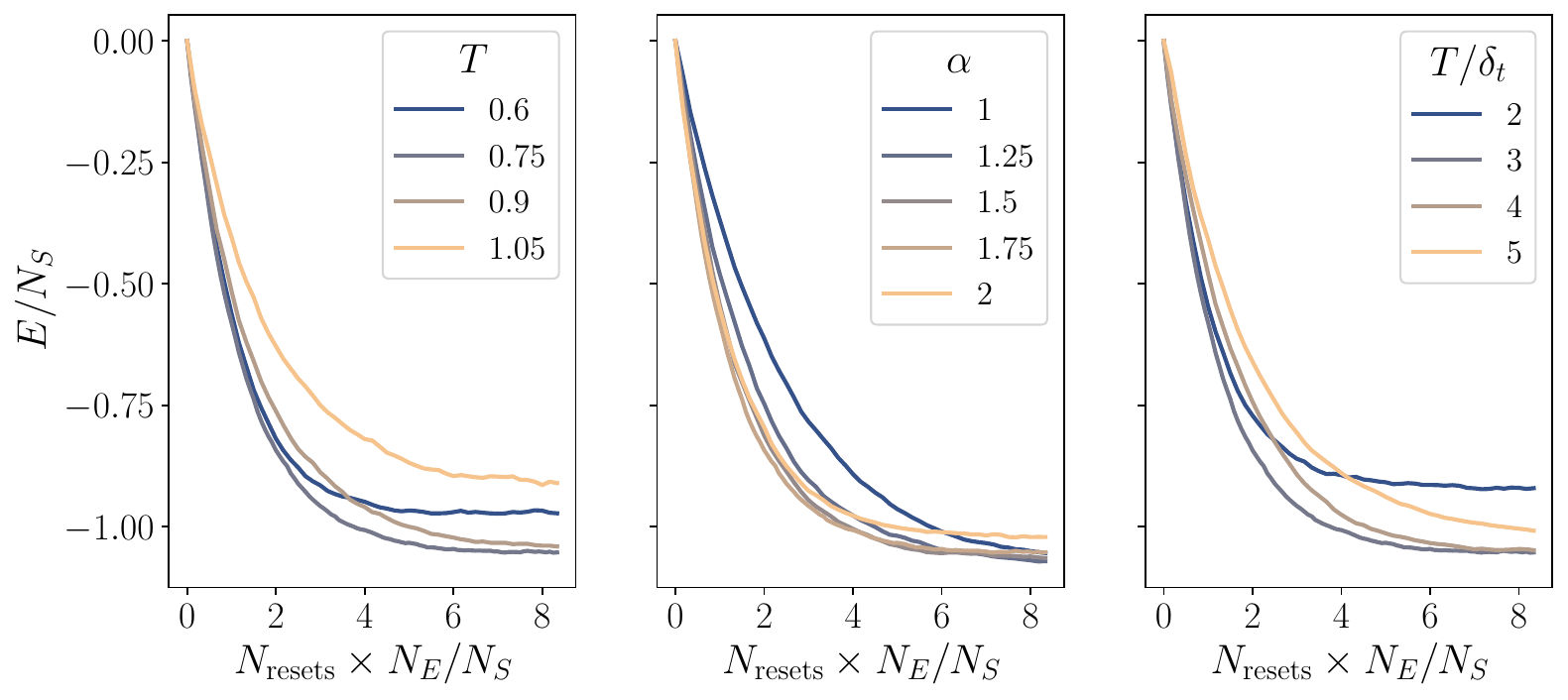}
    \caption{The convergence in the energy density on a $N_S=12$ kagome lattice with OBCs. The parameters are swept across those used in the implementation on {\tt ibm\_boston} that had $\{\delta_t, T,\alpha\} =\{0.25,0.75,1.75\}$.
    The left plot fixes $\alpha=1.75$ and the total number of Trotter steps $T/\delta_t =3$ while varying the total evolution time $T$.
    The center plot fixes $T=0.75$ and $\delta_t=0.25$ and varies the system-environment coupling $\alpha$.
    The right plot fixes $\alpha=1.75$ and $\delta_t=0.25$ and varies the number of Trotter steps.
    All plots are averaged over 1000 trajectories starting from a random initial product state and have $\sigma=\infty$, $\omega_{\text{max}}=8$, $\beta=\infty$.}
    \label{fig:paramSweep}
\end{figure}

The implementation of the dissipative quantum channel on {\tt ibm\_boston} attempts to balance the fixed point error of the noiseless steady state against the quantum resources required.
This amounts to a tuning of the parameters that specify the quantum channel.
We chose $\{\delta_t, T,\alpha\} =\{0.25,0.75,1.75\}$ based on the results of classical statevector simulations on a $N_S=12$ kagome lattice with OBCs.
The convergence in energy is shown across a parameter sweep in Fig.~\ref{fig:paramSweep}.
The parameters were chosen to converge to the steady state in as few reset cycles as possible, without significantly increasing the steady state energy.
Faster convergence increases the strength of the dissipative quantum channel which makes the prepared thermal state more resilient to device noise as discussed in Appendix~\ref{app:noiseEffect}.
The left plot shows that $T=0.75$ is optimal with a fixed budget of three Trotter steps.
The center plot shows that $\alpha=1.75$ is ideal for our simulations; any lower and the convergence is slower and any higher increases the fixed point energy.
The right plot shows the convergence with increasing number of Trotter steps but fixed $\delta_t=0.25$.
For this step size, three Trotter step leads to the fastest convergence to the steady state with the smallest steady state error.
Increasing the number of Trotter steps increases the steady state error due to increased Trotter error.
As shown in Table~\ref{tab:energies}, the steady state energy density of $E_{\text{s.s.}}^{(\text{Noiseless})}/N_S =  -1.05$ that is reached is still far from the exact ground state energy of $E_{\text{QMC}}/N_S=-1.40$.
Further decreasing the steady state error requires simultaneously decreasing $\delta_t$ while increasing $T$, as was done in the classical simulations in  Section~\ref{sec:classicalsims}.

\begin{figure}
    \centering
    \includegraphics[width=0.95\linewidth]{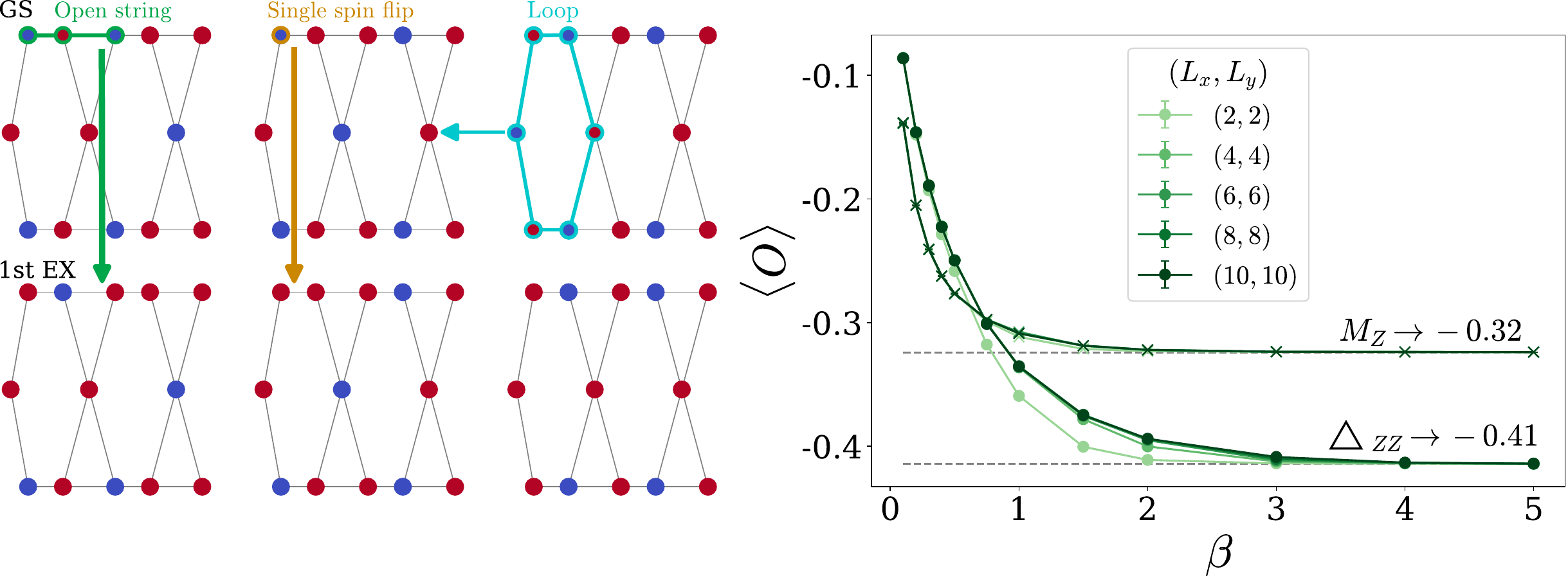}
    \caption{Left: All degenerate ground states and 3 of the degenerate first excited states of a classical Ising model with longitudinal field $g_z = 1.8$ on a small kagome-like lattice with OBCs. 
    Red sites denote spin up, while blue sites are spin down.  
    The degenerate ground states can be explored using spin-flips around closed loops of staggered spins which connect two configurations within the ice manifold.
    Transitioning from one of the ground states to one of the first excited states requires either a single spin flip or spin flips along a staggered, i.e. (up, down, up, down,...), open string. 
    Right: QMC results for the scaling of the magnetization density $M_Z$ and the connected $ZZ$ correlations around a triangle $\triangle_{ZZ}$, defined in Eq.~\eqref{eq:TriZZ}, for $g_x=0.5,g_z=2$ and varying kagome lattice geometries with PBCs. }
    \label{fig:SpinConfigs}
\end{figure}
\begin{figure}
    \centering
    \includegraphics[width=0.95\linewidth]{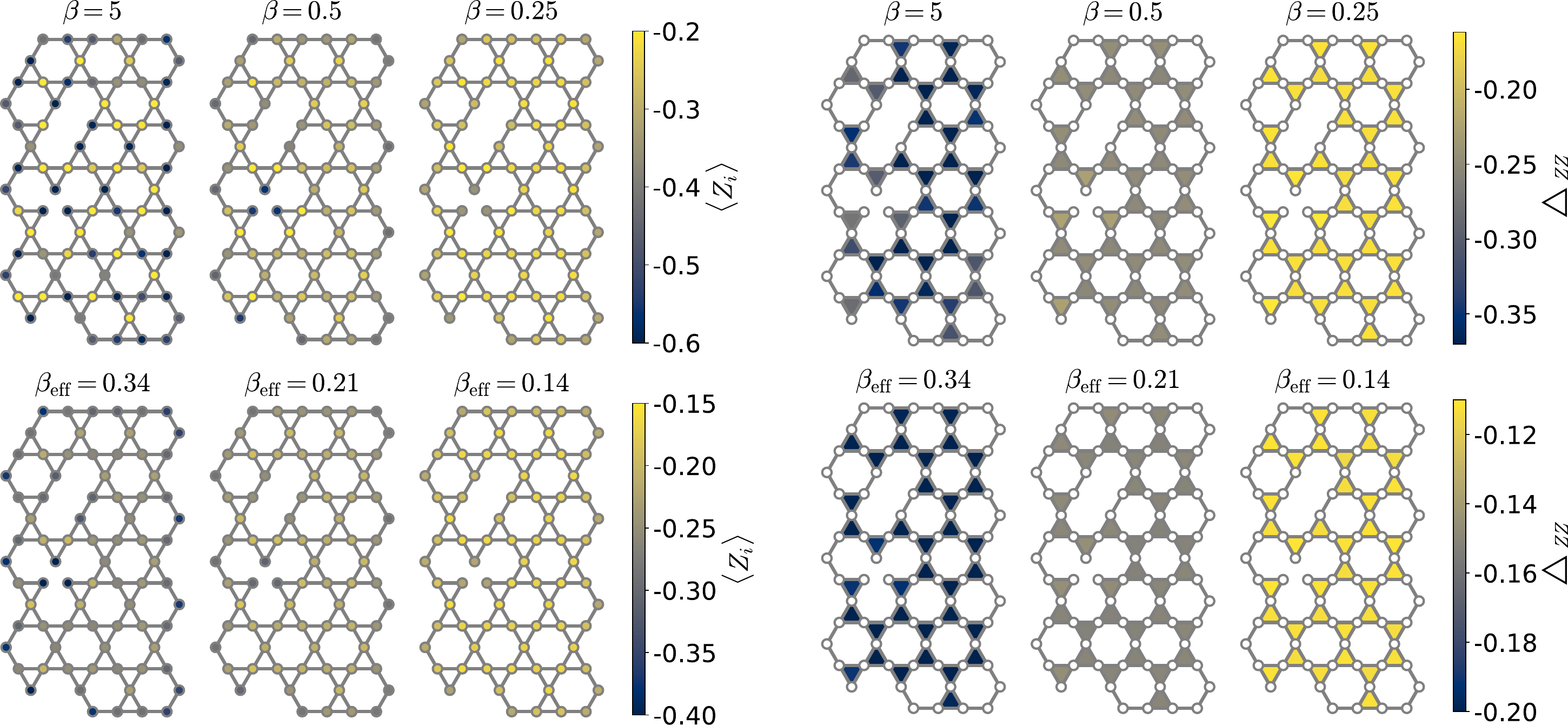}
    \caption{
    The spatial distribution of the magnetization (left) and connected triangle $ZZ$ correlations (right) from QMC simulations at different temperatures.
    The first row is for the values of $\beta$ used on {\tt ibm\_boston}, where we have substituted $\beta = 5$ for $\beta=\infty$. 
    The second row gives the observables at the effective temperatures $\beta_{\text{eff}}$ chosen to reproduce the energy measured on ${\tt ibm\_boston}$.}
    \label{fig:QMC_QC_comp_spatially}
\end{figure}

\section{Quantum Monte Carlo simulations} \label{app:QMC} 
\noindent
We use QMC simulations to obtain highly accurate determinations of observables when exact methods are not viable due to the exponential growth of the Hilbert space. 
In particular, QMC provides the only comparison to the results obtained from {\tt ibm\_boston} on the $N_S=79$ kagome lattice in Section~\ref{sec:qsims}.
In QMC, the two-dimensional lattice is mapped to a $(2+1)D$ system using the Suzuki-Trotter decomposition, where the additional dimension is imaginary time~\cite{RelationshipDDimensionalQuantal}. In order to use Monte Carlo updates from classical ice systems, we stay in discrete imaginary time with a Trotter step size of $\delta\tau = 0.02/g_x$~\cite{Narasimhan:2023inw}. 
Several different update rules are used to accommodate the geometric frustration and lattice defects on the $N_S=79$ lattice. 
Membrane updates are used to transition between the nearly degenerate states in the ice manifold.
The membranes are initialized from a loop obeying the ice-rules in one imaginary time slice and are then grown with a Wolff cluster-like algorithm in imaginary time~\cite{henryOrderbyDisorderQuantumCoulomb2014}. 
These updates remain self-balanced in the presence of arbitrary site defects. For bond defects between existing sites, the update requires that triangles are either completely intact or entirely removed, forbidding partially broken triangles. 
This is the case in the $N_S=79$ geometry chosen, see Appendix~\ref{app:bigKagome}.
Combining this with world line updates is sufficient for the AFIM with PBCs on a kagome lattice for temperatures $\beta \leq 10$~\cite{Narasimhan:2023inw,wang_tuning_2020}. 

We extend this previous result for our special case of OBCs and site defects. 
The necessary modifications to the update rules can be informed from looking at the structure of low-energy states in small systems.
An example of three degenerate ground states and three degenerate excited states are shown in Fig.~\ref{fig:SpinConfigs} for the classical Ising model with $g_x=0$.
The excited states can be reached from the degenerate ground states by either single spin flips or by flipping an open string of staggered spins. 
Therefore, we consider both single spin flips and open strings when applying QMC to larger lattices with OBCs.
These strings are constructed to be self-avoidant and of varying length. 
The strings are then grown in imaginary time. 
To make this construction satisfy classical detailed balance, the forward and backward probabilities need to be  explicitly calculated. 
This comes at a computational cost, which is not prohibitive for the system sizes studied in this work. 
The statistical errors due to a finite number of samples that is reported on the quantities in Section~\ref{sec:qsims} are obtained using the blocking technique from Refs.~\cite{flyvbjergErrorEstimatesAverages1989,defeverRsdefeverBlock_average2025a}.

For the $N_S =79$ lattice, where the ground state energy is not available from exact diagonalization, simulations with $\beta \leq 5$ are used to do a zero temperature extrapolation and get an estimate of the ground state energy.
This temperature is far from the quantum spin liquid regime that may arise in the thermodynamic limit at very low temperatures, the simulation of which remains an open challenge.
However, due to boundary effects, it is unlikely that such a phase exists on the $N_S=79$ lattices we use.
To mitigate finite-size and boundary effects, we examine the magnetization density and connected $\triangle_{ZZ}$ correlator, defined in Eq.~\eqref{eq:TriZZ},  on kagome lattices of varying sizes and PBCs.
This is shown across a sweep of temperatures in Fig.~\ref{fig:SpinConfigs}.
The observables have converged to their low-temperature values for $\beta>3$, and to the thermodynamic limit for $(L_x,L_y) > (4,4)$.
The extrapolated values of $M_Z=-0.32$ and $\triangle_{ZZ}=-0.41$ are quite close to the prediction of classical kagome spin ice of $M_Z=-1/3$ and $\triangle_{ZZ}=-4/9$.
Deviations are due to the non-zero longitudinal field, $g_x=0.5$.

In Section~\ref{sec:qsims}, an effective inverse-temperature $\beta_\text{eff}$ was reported.
The energy at the effective temperature is defined to be equal to the steady state energy measured on {\tt ibm\_boston}, see Eq.~\eqref{eq:Beff}. 
This is computed by performing a bisection search over temperatures until a temperature is found that reproduces the desired energy.
The energy at each temperature is computed with QMC following previous work in the context of dynamical phase transitions~\cite{Katschke:2026nmj}. 
The uncertainty in $\beta_\text{eff}$ reflects the range over which the QMC thermal energy agrees with the experimental data within its error bars. 
Once the global energy is matched to an effective temperature, the local observables are computed in  the thermal ensemble.

Fig.~\ref{fig:QMC_QC_comp_spatially} shows the magnetization density and $\triangle_{ZZ}$ correlators evaluated on the $N_S=79$ lattice using QMC.
The top row is for the values of $\beta$ that were run on {\tt ibm\_boston}, and the second row is for $\beta_{\text{eff}}$.
Since QMC simulations cannot reach $\beta=\infty$, we instead perform simulations at $\beta=5$.
The magnetization for $\beta>1/3$ shows that spins near the boundaries are more polarized than those in the bulk.
This is due to them having fewer antiferromagnetic neighbors.
Increasing the temperature causes the magnetization to trend to zero and become uniform across the lattice.
The $\triangle_{ZZ}$ correlators notably show significant less sensitivity to boundaries, and vary only slightly.
This observable is more sensitive to temperature fluctuations as it is a measure of geometric frustration. 
This is reflected in its sharper temperature dependence as shown in Fig.~\ref{fig:IBM_results_Energy_nE}{\bf b}.

\twocolumngrid
\bibliography{bibtemp}

\end{document}